\documentclass[manuscript]{acmart}

\usepackage{listings}
\usepackage{xcolor}
\usepackage{color}
\usepackage{multirow}
\usepackage{fontawesome5}
\usepackage{tikz}
\usepackage{pgfplots}
\usepackage{tabularray}
\usepackage{colortbl}
\usetikzlibrary{arrows.meta, positioning, calc}

\tikzstyle{box} = [rectangle, minimum width=4.5cm, minimum height=1.2cm, text centered, draw=black]
\tikzstyle{data} = [box, fill=blue!20]
\tikzstyle{template} = [box, fill=orange!30]
\tikzstyle{engine} = [box, fill=gray!30, align=center]
\tikzstyle{output} = [box, fill=green!30]
\tikzstyle{arrow} = [thick, ->, >=stealth]

\definecolor{codegreen}{rgb}{0,0.6,0}
\definecolor{codegray}{rgb}{0.5,0.5,0.5}
\definecolor{codepurple}{rgb}{0.58,0,0.82}
\definecolor{backcolour}{rgb}{0.95,0.95,0.92}
\definecolor{lightgray}{rgb}{.9,.9,.9}
\definecolor{darkgray}{rgb}{.4,.4,.4}
\definecolor{purple}{rgb}{0.65, 0.12, 0.82}
\definecolor{darkred}{rgb}{0.6,0.0,0.0}
\definecolor{darkgreen}{rgb}{0,0.50,0}
\definecolor{lightblue}{rgb}{0.0,0.42,0.91}
\definecolor{orange}{rgb}{0.99,0.48,0.13}
\definecolor{grass}{rgb}{0.18,0.80,0.18}
\definecolor{pink}{rgb}{0.97,0.15,0.45}
\definecolor{mygreen}{rgb}{0,0.6,0}
\definecolor{mygray}{rgb}{0.5,0.5,0.5}
\definecolor{mymauve}{rgb}{0.58,0,0.82}
\definecolor{Concrete}{rgb}{0.949,0.949,0.949}

\lstdefinelanguage{JavaScript}{
  keywords={typeof, new, true, false, catch, function, return, null, catch, switch, var, if, in, while, do, else, case, break},
  keywordstyle=\color{blue}\bfseries,
  ndkeywords={class, export, boolean, throw, implements, import, this},
  ndkeywordstyle=\color{darkgray}\bfseries,
  identifierstyle=\color{black},
    sensitive=false,
  comment=[l]{//},
  morecomment=[s]{/*}{*/},
    commentstyle=\color{purple}\ttfamily,
  stringstyle=\color{red}\ttfamily,
  morestring=[b]',
  morestring=[b]"
}

\AtBeginDocument{%
  }

\setcopyright{acmcopyright}
\copyrightyear{2025}
\acmYear{2025}
\acmDOI{XXXXXXX.XXXXXXX}

\begin{document}

\title[An Assessment of the Overlooked Dangers of Template Engines]{An Assessment of the Overlooked Dangers of Template Engines}

\author{Lorenzo Pisu}
\email{lorenzo.pisu@unica.it}
\orcid{0009-0001-0129-1976}
\affiliation{%
  \institution{University Of Cagliari}
  \streetaddress{Piazza D'Armi}
  \city{Cagliari}
  \country{Italy}
  \postcode{09123}
}

\author{Davide Maiorca}
\affiliation{%
  \institution{University Of Cagliari}
  \streetaddress{Piazza D'Armi}
  \city{Cagliari}
  \country{Italy}
  \postcode{09123}
  }
\email{davide.maiorca@unica.it}
\orcid{0000-0003-2640-4663}

\author{Giorgio Giacinto}
\affiliation{%
  \institution{University Of Cagliari}
  \streetaddress{Piazza D'Armi}
  \city{Cagliari}
  \country{Italy}
  \postcode{09123}
  }
\affiliation{%
  \institution{National Interuniversity Consortium for Informatics}
  \streetaddress{Piazza D'Armi}
  \city{Cagliari}
  \country{Italy}
  \postcode{09123}
  }
\email{giacinto@unica.it}
\orcid{0000-0002-5759-3017}

\renewcommand{\shortauthors}{Pisu L., Maiorca D., Giacinto G.}

\begin{abstract}
Template engines play a pivotal role in modern web application development by enabling the dynamic rendering of content, products, and user interfaces. Today, they are essential for any website that handles dynamic data, from e-commerce to social media. However, their widespread adoption also makes them attractive targets for attackers seeking to exploit vulnerabilities and gain unauthorized access to web servers.

This paper presents a comprehensive assessment of the risks associated with template engines, with a particular focus on the consequences of Server-Side Template Injection (SSTI) and the ease with which such vulnerabilities can escalate to Remote Code Execution (RCE), a critical security concern in web application development.
\end{abstract}




\maketitle

\section{Introduction}
\label{sec:intro}
Template engines~\cite{mardan2018template, parr2004enforcing} are crucial tools in both web development and other software applications, as they help separate the presentation layer from the application's logic. They allow developers to create templates or patterns for generating dynamic content, which can be rendered as HTML, XML, or other markup languages based on data or variables provided at runtime. Template engines are software components, typically provided as libraries or modules, that offer a set of functions to parse and manipulate strings or files according to predefined syntactic rules. Moreover, template engines usually apply tokenization, breaking strings or files into structured representations. This process allows binding data to placeholders, applying transformations, and executing conditional logic and loops. The main scenario in which template engines are used is when we want to serve dynamic content on a website. For example, providing a dashboard that contains information that changes based on the user visiting it. A concrete use case can be an e-commerce website that displays product listings. These listings often involve dynamic content, such as product names, prices, availability, and user-specific recommendations, making template engines essential for rendering these pages with up-to-date information for each product. Countless examples of modern websites require template engines, making this technology widely adopted. While template engines provide many benefits, they also come with potential security pitfalls that developers should be aware of. Server-Side Template Injection (SSTI) is the main vulnerability linked to template engines. SSTI is an injection vulnerability in the OWASP top 10 vulnerabilities list~\cite{owasptop10injection}, and its impact can be potentially critical. The worst consequence of SSTI exploitation is achieving Remote Code Execution (RCE)~\cite{holm2012success, biswas2018study}, allowing attackers to take control of the entire target server. RCE is a common consequence of SSTI because most template engines can be used by attackers as paths to execute arbitrary code. However,
SSTI can lead to many other potential consequences, such as:
\begin{itemize}
    \item \textbf{Information Disclosure}: SSTI can expose server-side configuration files, source code, and other sensitive information that can aid attackers in further exploiting the system.
    \item \textbf{Unauthorized Access}: with SSTI, attackers can gain unauthorized access to restricted areas of the application or server, potentially taking control of admin panels or other privileged functionalities.
    \item \textbf{DoS Attacks}: SSTI can be leveraged to launch Denial of Service (DoS) attacks on the server or related systems, disrupting services for legitimate users.
    \item \textbf{Cross-Site Scripting}: XSS is a common symptom of SSTI that attackers can exploit to steal sensitive information from legitimate users.
\end{itemize}

Although these consequences of SSTI are severe, RCE represents the most critical threat, making it the central focus of this paper. To address this concern, we conduct a comprehensive analysis of widely adopted template engines, examining their susceptibility to RCE when exploited through SSTI vulnerabilities. By concentrating on this specific threat, our goal is to deepen the understanding of how template engine design and behavior can expose systems to RCE, thereby contributing to the development of more secure web applications and server environments.

Current research does not sufficiently address the issue of RCE in template engines despite the prevalence of numerous examples in real-world scenarios. In recent years, the focus of research has predominantly centered on the offensive and defensive aspects of SSTI, with the development of tools and strategies aimed at evading protections. However, the problem of RCE in template engines has largely been overlooked, lacking tools for assessment and isolation. Future research in this domain can provide ways to perform RCE detection and prevention by creating automated tools specifically designed for identifying RCE in template engines. Moreover, exploring new methods to fortify defenses against RCE is crucial, given the current susceptibility of most template engines. Our assessment aims to show the existing gaps in research concerning SSTI, RCE, and the overall security of template engines.

Furthermore, we will show the main tools used for SSTI detection, analyzing their functions and their main differences. We conduct a literature review on SSTI, as it offers interesting ideas on open research problems in this topic. Moreover, we will present details on the wide adoption and impact of template engines through the years, focusing on how widely they are used and how many vulnerabilities have been found since the discovery of SSTI.

We begin our assessment by offering insights into the utilization of template engines, showing how widely adopted this technology is. Our initial step involves querying GitHub to retrieve data on the number of repositories associated with template engines, providing a comprehensive overview of their prevalence over time. Additionally, we explore real-world instances of SSTI and Common Vulnerabilities and Exposures (CVEs)~\cite{cve} reported in the past. Leveraging the bug bounty platform HackerOne~\cite{hackerone}, we gather reports pertaining to SSTI, while utilizing the CVE search engine to identify frameworks and applications that have documented vulnerabilities in relation to SSTI. This multifaceted approach allows us to show both the broad adoption and specific security issues surrounding template engines.

Then, we show template engines' essential features and practical code examples of how they are used to render dynamic content. We also investigate the underlying mechanisms that make template engines prone to allowing RCE, examining various scenarios. Additionally, we will present a methodology that we developed to analyze RCE attacks and defenses in template engines and the results obtained by applying this methodology to 34 template engines in eight programming languages. From the results, we show four categories of RCE paths present in the template engine we analyzed. Furthermore, we explore mitigation strategies and best practices for developers to protect against RCE attacks when utilizing template engines. We will categorize these strategies into four methods we encountered during our analysis. We will also explore other strategies related to SSTI protection involving input validation and secure template engine configurations. Our assessment also highlights the importance of maintaining up-to-date dependencies, as some template engines evolve to address security issues. We discuss the role of the developer and the template engine community in reducing the risk of RCE vulnerabilities and fostering a more secure web development environment. 

In the following, we summarize the key contributions of this paper:

\begin{itemize}
\item We present and contextualize the usage and prevalence of template engines in real-world applications, including examples of critical SSTI vulnerabilities and associated CVEs. We also introduce the template engines selected for our study, highlighting their general characteristics, popularity (e.g., GitHub stars), and, where available, monthly download statistics.

\item We offer a comprehensive overview of SSTI vulnerabilities by identifying the template engine behaviors that enable RCE, and examining the different scenarios in which these vulnerabilities arise.

\item We propose a taxonomy of RCE types encountered in template engines, as well as a classification of the prevention mechanisms implemented to mitigate such risks.

\item We review existing literature on SSTI, covering attack methodologies, sandbox escape techniques, and defense strategies. We also identify and discuss significant research gaps that remain unaddressed.

\item We analyze the main SSTI detection tools currently available, outlining their capabilities, limitations, and practical implications for vulnerability identification.

\item We propose a systematic methodology for assessing the security posture of template engines, from the feasibility of achieving RCE to the presence and effectiveness of implemented protections.

\item We apply our methodology to evaluate 34 template engines across eight different programming languages, offering a comprehensive, comparative analysis intended to inform both researchers and practitioners.

\item We present detailed case studies on specific template engines that yielded particularly notable results during our evaluation.
\end{itemize}

The rest of the article is structured as follows: Section~\ref{sec:SSTIinTheWild} shows how template engines and SSTI have evolved in these years, analyzing usage, CVEs, and real-world bugs. Section~\ref{sec:teAndSSTI} provides a general overview of how template engines work, how SSTI arises, its consequences, and why arbitrary code execution is a common problem. Section~\ref{sec:RCEinTE} focuses on RCE in template engines, explaining the categories, the defenses, and their limitations. Section~\ref{sec:literature} summarizes the most important works on SSTI. Section~\ref{sec:detectionTools} provides an overview of the existing SSTI detection tools, describing their features and limitations.  Section~\ref{sec:templateAnalysis} presents the results of our analysis on 34 template engines, as well as the methodology we adopted to carry it out. Section~\ref{sec:examples} focuses on five template engines we encountered during our analysis,  providing a detailed analysis. Section~\ref{sec:lessons_learned} provides a summary of the main takeaways of our work. Section~\ref{sec:conclusion} concludes the article by highlighting possible future research directions on the problem of SSTI and RCE in template engines.

\section{Template Engines And SSTI In The Wild}
\label{sec:SSTIinTheWild}
\begin{figure}[H]
\includegraphics[width=\textwidth]{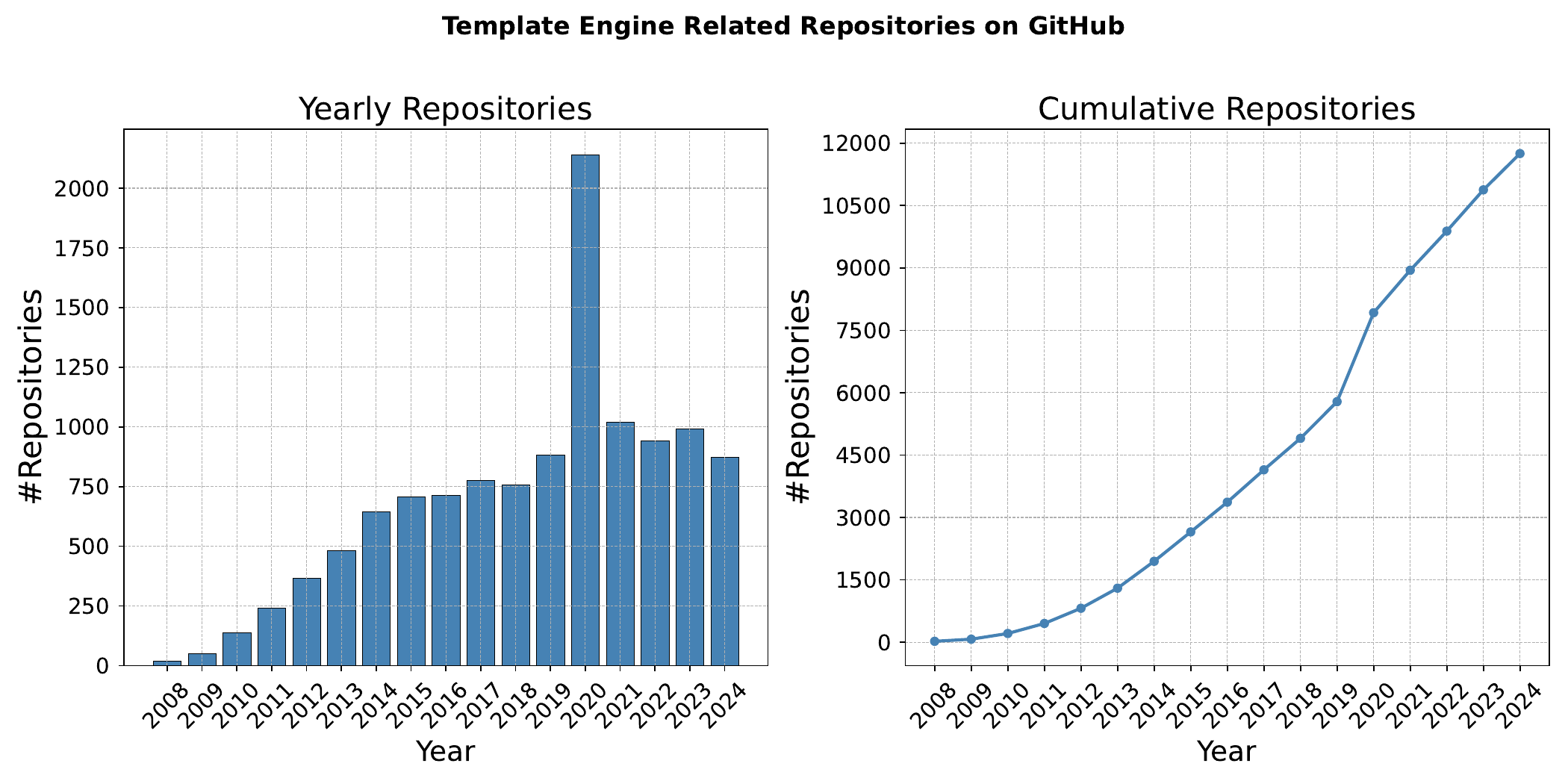}
\caption{These plots show how many repositories respond to the \emph{template engine} search query on GitHub. On the left, we show the number of repositories per year based on the creation date. On the right, we represent the cumulative sum. A notable increase in the number of repositories was observed in 2020. While the exact causes remain unclear, this trend coincides with the onset of the COVID-19 pandemic, which has been widely associated with shifts in developer behavior, including increased open-source contributions and remote software development activities~\cite{wang2021understanding}.}
\label{fig:templatesgithub}

\end{figure}

Before diving into the specifics of how template engines work, we gather and show statistics related to this technology adoption, prevalence, and impact in the modern web application landscape. This section is divided into four parts: the first is centered around template engine usage, and the other three focus on the problems related to this technology, namely SSTI and RCE. 
Figure~\ref{fig:templatesgithub} shows how many repositories we find searching \texttt{template engine} on GitHub. We divide the results by the creation date of the repositories, considering years from $2008$ to $2024$. We present each year's repository count and a cumulative total. While not every repository in the results is a template engine, it is reasonable to infer a consistent annual rise in the number of template engines. This trend signifies that, despite the abundance of existing template engines, new ones are continually emerging. This is due to the nature of template engines, which can vary in the way they function and in how efficient and fast they are.

\subsection{Template Engines Usage}
\label{sec:SSTIinTheWild:templateUsage}

Since each programming language needs its own set of template engine libraries, there is a wide variety from which programmers can choose depending on their needs. Therefore, when considering the present and future use of template engines, we must also account for the development of new engines tailored to emerging programming languages.

Most template engines are open-source and are often associated with popular frameworks such as Flask~\cite{flaskdocs}, which uses Jinja2~\cite{jinja2docs}, or Laravel~\cite{laraveldocs}, which has its own engine called Blade~\cite{bladedocs}. To understand their widespread adoption, we can analyze highly popular GitHub repositories (based on their star ratings) and the monthly download statistics of frameworks built on these templates. There are also numerous lesser-known or proprietary template engines, indicating that the use of such technology is even more extensive. Table \ref{tab:popularity} presents an overview of the template engines under analysis and their susceptibility to RCE. We observe that many of the most widely used template engines permit RCE, a concern that will be further examined in Section~\ref{sec:RCEinTE}. This suggests that RCE is not usually taken into account when selecting a template engine. In fact, the popularity of web frameworks such as Flask, Vue, and Laravel indicates that the choice of template engine is often not prioritized, with default or user-friendly engines being the most commonly adopted.

\begin{table*}[h]
\begin{tabular}{cl|l|ccc|c} 
\toprule
\multirow{2}{*}{\textbf{Language}}    & \multirow{2}{*}{\textbf{Template Engine}}           & \multirow{2}{*}{\textbf{Ref.}}                                 & \multicolumn{3}{c|}{\textbf{Popularity}}                                                                                            & \multirow{2}{*}{\textbf{Allows RCE}}                \\ 
\cmidrule(l){4-6}
                                      &                                                     &                                                                & \faStar                                   & \faCodeBranch                             & \faDownload                                 &                                                     \\ 
\midrule
\multirow{9}{*}{\textbf{Python}}      & Django                                              & \cite{repoDjango}                                              & 83.3k                                     & 32.5k                                     & 21.3M                                       & \faTimes                                            \\
                                      & {\cellcolor[rgb]{0.949,0.949,0.949}}Tornado         & {\cellcolor[rgb]{0.949,0.949,0.949}}\cite{repoTornado}         & {\cellcolor[rgb]{0.949,0.949,0.949}}21.9k & {\cellcolor[rgb]{0.949,0.949,0.949}}5.5k  & {\cellcolor[rgb]{0.949,0.949,0.949}}55.8M   & {\cellcolor[rgb]{0.949,0.949,0.949}}\faCheck        \\
                                      & Jinja2                                              & \cite{repoJinja}                                               & 10.8k                                     & 1.6k                                      & 293.7M                                      & \faCheck                                            \\
                                      & {\cellcolor[rgb]{0.949,0.949,0.949}}web2py          & {\cellcolor[rgb]{0.949,0.949,0.949}}\cite{repoWeb2py}          & {\cellcolor[rgb]{0.949,0.949,0.949}}2.1k  & {\cellcolor[rgb]{0.949,0.949,0.949}}907   & {\cellcolor[rgb]{0.949,0.949,0.949}}305     & {\cellcolor[rgb]{0.949,0.949,0.949}}\faCheck        \\
                                      & Mako                                                & \cite{repoMako}                                                & 386                                       & 64                                        & 52.9M                                       & \faCheck                                            \\
                                      & {\cellcolor[rgb]{0.949,0.949,0.949}}Chameleon       & {\cellcolor[rgb]{0.949,0.949,0.949}}\cite{repoChameleon}       & {\cellcolor[rgb]{0.949,0.949,0.949}}182   & {\cellcolor[rgb]{0.949,0.949,0.949}}65    & {\cellcolor[rgb]{0.949,0.949,0.949}}101.5k  & {\cellcolor[rgb]{0.949,0.949,0.949}}\faCheck        \\
                                      & Cheetah3                                            & \cite{repoCheetah3}                                            & 147                                       & 37                                        & 696.1k                                      & \faCheck                                            \\
                                      & {\cellcolor[rgb]{0.949,0.949,0.949}}Genshi          & {\cellcolor[rgb]{0.949,0.949,0.949}}\cite{repoGenshi}          & {\cellcolor[rgb]{0.949,0.949,0.949}}92    & {\cellcolor[rgb]{0.949,0.949,0.949}}36    & {\cellcolor[rgb]{0.949,0.949,0.949}}136.3k  & {\cellcolor[rgb]{0.949,0.949,0.949}}\faCheck        \\
                                      & Pyratemp                                            & \cite{pyratempdocs}                                            & -                                         & -                                         & 3.5k                                        & \faTimes                                            \\ 
\midrule
\multirow{3}{*}{\textbf{PHP}}         & {\cellcolor[rgb]{0.949,0.949,0.949}}Laravel         & {\cellcolor[rgb]{0.949,0.949,0.949}}\cite{repoLaravel}         & {\cellcolor[rgb]{0.949,0.949,0.949}}80.7k & {\cellcolor[rgb]{0.949,0.949,0.949}}24.3k & {\cellcolor[rgb]{0.949,0.949,0.949}}7.5M    & {\cellcolor[rgb]{0.949,0.949,0.949}}\faCheck        \\
                                      & Twig                                                & \cite{repoTwig}                                                & 8.2k                                      & 1.2k                                      & 5M                                          & \faExclamation                                      \\
                                      & {\cellcolor[rgb]{0.949,0.949,0.949}}Smarty          & {\cellcolor[rgb]{0.949,0.949,0.949}}\cite{repoSmarty}          & {\cellcolor[rgb]{0.949,0.949,0.949}}2.3k  & {\cellcolor[rgb]{0.949,0.949,0.949}}715   & {\cellcolor[rgb]{0.949,0.949,0.949}}468.9 k & {\cellcolor[rgb]{0.949,0.949,0.949}}\faExclamation  \\ 
\midrule
\multirow{11}{*}{\textbf{JavaScript}} & Vue                                                 & \cite{repoVue}                                                 & 208.7k                                    & 33.7k                                     & 24.3M                                       & \faCheck                                            \\
                                      & {\cellcolor[rgb]{0.949,0.949,0.949}}Pug             & {\cellcolor[rgb]{0.949,0.949,0.949}}\cite{repoPug}             & {\cellcolor[rgb]{0.949,0.949,0.949}}21.8k & {\cellcolor[rgb]{0.949,0.949,0.949}}1.9k  & {\cellcolor[rgb]{0.949,0.949,0.949}}6.8M    & {\cellcolor[rgb]{0.949,0.949,0.949}}\faCheck        \\
                                      & Handlebars                                          & \cite{repoHandlebars.js}                                       & 18.2k                                     & 2k                                        & 67.7M                                       & \faExclamation                                      \\
                                      & {\cellcolor[rgb]{0.949,0.949,0.949}}Marko           & {\cellcolor[rgb]{0.949,0.949,0.949}}\cite{repoMarko}           & {\cellcolor[rgb]{0.949,0.949,0.949}}13.5k & {\cellcolor[rgb]{0.949,0.949,0.949}}650   & {\cellcolor[rgb]{0.949,0.949,0.949}}46.2k   & {\cellcolor[rgb]{0.949,0.949,0.949}}\faCheck        \\
                                      & Nunjucks                                            & \cite{repoNunjucks}                                            & 8.6k                                      & 645                                       & 3.9M                                        & \faCheck                                            \\
                                      & {\cellcolor[rgb]{0.949,0.949,0.949}}EJS             & {\cellcolor[rgb]{0.949,0.949,0.949}}\cite{repoEjs}             & {\cellcolor[rgb]{0.949,0.949,0.949}}7.9k  & {\cellcolor[rgb]{0.949,0.949,0.949}}852   & {\cellcolor[rgb]{0.949,0.949,0.949}}81.1M   & {\cellcolor[rgb]{0.949,0.949,0.949}}\faCheck        \\
                                      & doT                                                 & \cite{repoDot}                                                 & 5k                                        & 1k                                        & 1.9M                                        & \faCheck                                            \\
                                      & {\cellcolor[rgb]{0.949,0.949,0.949}}Dust            & {\cellcolor[rgb]{0.949,0.949,0.949}}\cite{repoDustjs}          & {\cellcolor[rgb]{0.949,0.949,0.949}}2.9k  & {\cellcolor[rgb]{0.949,0.949,0.949}}475   & {\cellcolor[rgb]{0.949,0.949,0.949}}51.2k   & {\cellcolor[rgb]{0.949,0.949,0.949}}\faExclamation  \\
                                      & JsRender                                            & \cite{repoJsrender}                                            & 2.6k                                      & 339                                       & 63.5k                                       & \faCheck                                            \\
                                      & {\cellcolor[rgb]{0.949,0.949,0.949}}Template7       & {\cellcolor[rgb]{0.949,0.949,0.949}}\cite{repoTemplate7}       & {\cellcolor[rgb]{0.949,0.949,0.949}}659   & {\cellcolor[rgb]{0.949,0.949,0.949}}165   & {\cellcolor[rgb]{0.949,0.949,0.949}}6k      & {\cellcolor[rgb]{0.949,0.949,0.949}}\faCheck        \\
                                      & SquirrellyJS                                        & \cite{repoSquirrelly}                                          & 657                                       & 83                                        & 105.5k                                      & \faCheck                                            \\ 
\midrule
\multirow{5}{*}{\textbf{Java}}        & {\cellcolor[rgb]{0.949,0.949,0.949}}Thymeleaf       & {\cellcolor[rgb]{0.949,0.949,0.949}}\cite{repoThymeleaf}       & {\cellcolor[rgb]{0.949,0.949,0.949}}2.8k  & {\cellcolor[rgb]{0.949,0.949,0.949}}511   & {\cellcolor[rgb]{0.949,0.949,0.949}}-       & {\cellcolor[rgb]{0.949,0.949,0.949}}\faExclamation  \\
                                      & Pebble                                              & \cite{repoPebble}                                              & 1.1k                                      & 170                                       & -                                           & \faExclamation                                      \\
                                      & {\cellcolor[rgb]{0.949,0.949,0.949}}FreeMarker      & {\cellcolor[rgb]{0.949,0.949,0.949}}\cite{repoFreemarker}      & {\cellcolor[rgb]{0.949,0.949,0.949}}1k    & {\cellcolor[rgb]{0.949,0.949,0.949}}271   & {\cellcolor[rgb]{0.949,0.949,0.949}}-       & {\cellcolor[rgb]{0.949,0.949,0.949}}\faCheck        \\
                                      & Jinjava                                             & \cite{repoJinjava}                                             & 725                                       & 170                                       & -                                           & \faExclamation                                      \\
                                      & {\cellcolor[rgb]{0.949,0.949,0.949}}Apache Velocity & {\cellcolor[rgb]{0.949,0.949,0.949}}\cite{repoVelocity-engine} & {\cellcolor[rgb]{0.949,0.949,0.949}}385   & {\cellcolor[rgb]{0.949,0.949,0.949}}134   & {\cellcolor[rgb]{0.949,0.949,0.949}}-       & {\cellcolor[rgb]{0.949,0.949,0.949}}\faCheck        \\ 
\midrule
\multirow{2}{*}{\textbf{Ruby}}        & Slim                                                & \cite{repoSlim}                                                & 5.3k                                      & 503                                       & -                                           & \faCheck                                            \\
                                      & {\cellcolor[rgb]{0.949,0.949,0.949}}ERB             & {\cellcolor[rgb]{0.949,0.949,0.949}}\cite{repoErb}             & {\cellcolor[rgb]{0.949,0.949,0.949}}137   & {\cellcolor[rgb]{0.949,0.949,0.949}}19    & {\cellcolor[rgb]{0.949,0.949,0.949}}-       & {\cellcolor[rgb]{0.949,0.949,0.949}}\faCheck        \\ 
\midrule
\textbf{Go}                           & html/template                                       & \cite{goenginedocs}                                            & -                                         & -                                         & -                                           & \faTimes                                            \\ 
\midrule
\textbf{Perl}                         & {\cellcolor[rgb]{0.949,0.949,0.949}}Mojolicious     & {\cellcolor[rgb]{0.949,0.949,0.949}}\cite{repoMojo}            & {\cellcolor[rgb]{0.949,0.949,0.949}}2.6k  & {\cellcolor[rgb]{0.949,0.949,0.949}}583   & {\cellcolor[rgb]{0.949,0.949,0.949}}-       & {\cellcolor[rgb]{0.949,0.949,0.949}}\faCheck        \\ 
\midrule
\multirow{2}{*}{\textbf{.NET}}        & ASP                                                 & \cite{repoAspnetcore}                                          & 36.5k                                     & 10.3k                                     & -                                           & \faCheck                                            \\
                                      & {\cellcolor[rgb]{0.949,0.949,0.949}}Razor           & {\cellcolor[rgb]{0.949,0.949,0.949}}\cite{repoRazor}           & {\cellcolor[rgb]{0.949,0.949,0.949}}544   & {\cellcolor[rgb]{0.949,0.949,0.949}}203   & {\cellcolor[rgb]{0.949,0.949,0.949}}-       & {\cellcolor[rgb]{0.949,0.949,0.949}}\faCheck        \\
\bottomrule
\end{tabular}
\caption*{Legend:\textbf{\faStar=GitHub Stars; \faCodeBranch=GitHub Forks; \faDownload=Monthly Downloads; \faTimes=Does not allow RCE; \faExclamation=Allows or allowed RCE but also has protections; \faCheck=Allows RCE}}
\caption{Overview of the template engines under analysis along with their popularity and if they allow RCE. The Monthly downloads metric (\faDownload) was taken from pystats (for Python), packagist (for PHP), and NPM (for JavaScript).}
\label{tab:popularity}
\end{table*}

\begin{table}[]
\begin{tblr}{
  row{3} = {Concrete},
  row{5} = {Concrete},
  row{7} = {Concrete},
  column{odd} = {c},
  column{2} = {c},
  column{4} = {c},
  hline{1,9} = {-}{0.08em},
  hline{2} = {-}{0.05em},
}
\textbf{Report ID} & \textbf{Year} & \textbf{Keywords}        & \textbf{Reported to} & \textbf{Engine} & \textbf{Bounty (\$)} \\
\textbf{125980}    & 2016          & RCE, mail                & Uber                 & Jinja2          & 10,000               \\
\textbf{301406}    & 2017          & LFI, Requires privileges & Ubiquiti Inc.        & Twig            & 1,000+               \\
\textbf{423541}    & 2018          & RCE, mail                & Shopify              & Handlebars      & 10,000               \\
\textbf{536130}    & 2019          & RCE, CVE-2019-3396       & Mail.ru              & Velocity        & 2,000                \\
\textbf{1537543}   & 2022          & RCE, CVE-2022-22954      & U.S. Dept Of Defense & FreeMarker      & -                    \\
\textbf{1928279}   & 2023          & Ruby                     & GitHub Security Lab  & ERB, Slim       & 2,300                
\end{tblr}
\caption{Reports on HackerOne relative to SSTI (Some bounties have a plus(+) sign indicating that the author of the report received a bonus bounty that was not disclosed).}
\label{tab:hackeronereports} 
\end{table}

\subsection{SSTI Vulnerabilities in the Wild}
\label{sec:SSTIinTheWild:cves}
This section focuses on real-world applications that have been found vulnerable to SSTI and the impact of such findings on companies and agencies. In the following subsections, we will also show that many CVEs have been found about SSTI in popular frameworks and applications that use template engines.

The severity of SSTI (Server-Side Template Injection) in real-world websites and frameworks is confirmed by bug bounty reports. Table \ref{tab:hackeronereports} shows six SSTI vulnerability reports. The column \emph{reported to} shows that big companies and agencies can also have SSTI vulnerabilities. Moreover, the bounty payouts to the authors of those reports show that companies value SSTI as a critical vulnerability. The keywords demonstrate that SSTI often leads to RCE altogether, thus being critical for companies to fix. Furthermore, the constant presence of the vulnerability throughout the years shows that SSTI is still a problem that needs to be addressed properly in real-world applications.
SSTI corresponds to CWE-1336 (Improper Neutralization of Special Elements Used in a Template Engine), which is categorized under the broader CWE-94 (Code Injection). Despite having a precise CWE categorization, SSTI often appears under other CWE IDs, making it harder to precisely estimate the number of CVEs related to this vulnerability. Nevertheless, we have considered both CVEs categorized under CWE-1336 and searched for \emph{template injection}. Table \ref{tab:cves} shows 10 CVEs that feature SSTI vulnerabilities. The base score highlights how critical these vulnerabilities are when they arise and how RCE is often a direct consequence of SSTI. Besides, we can notice that the engines behind these CVEs are often the same (Twig, Velocity, FreeMarker), probably due to their popularity and language usage (Java and PHP). 
\begin{table}[]
\begin{tblr}{
  row{3} = {Concrete},
  row{5} = {Concrete},
  row{7} = {Concrete},
  row{9} = {Concrete},
  row{11} = {Concrete},
  column{odd} = {c},
  column{2} = {c},
  hline{1,13} = {-}{0.08em},
  hline{2} = {-}{0.05em},
}
\textbf{Vulnerability}  & \textbf{Base Score} & \textbf{Keywords}                        & \textbf{Engine} \\
\textbf{CVE-2017-16783} & 9.8                 & RCE, CMS Made Simple,                    & Smarty          \\
\textbf{CVE-2018-20465} & 7.2                 & Information disclosure, Authenticated    & Twig            \\
\textbf{CVE-2019-3396}  & 10                  & RCE                                      & Velocity        \\
\textbf{CVE-2019-19999} & 7.2                 & Misconfiguration                         & FreeMarker      \\
\textbf{CVE-2020-1961}  & 9.8                 & RCE, Apache Syncope                      & JEXL            \\
\textbf{CVE-2020-4027}  & 6.5                 & RCE, Requires Privileges                 & Velocity        \\
\textbf{CVE-2020-12790} & 7.5                 & Information disclosure, CraftCMS, plugin & Twig            \\
\textbf{CVE-2020-26282} & 10                  & RCE, BrowserUp Proxy                     & Java EL         \\
\textbf{CVE-2021-21244} & 10                  & RCE, OneDev                              & Java EL         \\
\textbf{CVE-2022-22954} & 10                  & RCE, VMware                              & FreeMarker            
\end{tblr}
 \caption{CVEs related to SSTI, we show the main keywords that appear in the CVE report and the template engine used by the vulnerable application.}
 \label{tab:cves}
\end{table}
 
\subsection{SSTI Scenarios}
\label{sec:SSTIinTheWild:scenarios}
There are two prominent instances of template engine applications encountered in real-world settings. Firstly, websites that utilize template engines for the dynamic rendering of HTML pages are the prevalent use case in contemporary websites. In this context, preventing SSTI is of crucial importance, and a comprehensive understanding of the template engine's functioning can mitigate the risks associated with RCE or SSTI vulnerabilities. 
Secondly, Content Management Systems (CMSs) often need to provide the user with templating functionalities. In this case, using an engine that does not allow RCE is essential. Another example is "Website as a Service" platforms like GitHub Pages or Netlify, which facilitate static web page hosting. For instance, GitHub Pages relies on \texttt{Jekyll}, a \texttt{Ruby} framework for building static websites, featuring the \texttt{Liquid} template engine among its components. Despite being used in popular frameworks, we did not analyze \texttt{Liquid} in detail in this work as it did not appear in any of the sources we used to select the template engines to analyze. Fortunately, while \texttt{Liquid} does not currently harbor RCE vulnerabilities, a hypothetical vulnerability could allow malevolent actors to exploit GitHub's build process to attain a reverse shell. This instance underlines the pivotal role of selecting a secure template engine for such applications. 

By categorizing template engine usage into these scenarios, we can underline the importance of selecting a template engine based on the application requirements. In a typical web application, SSTI must be avoided completely; therefore, selecting a non-RCE template engine is still important but less impactful. Vice versa, selecting a non-RCE engine is fundamental in a CMS or platform that allows its users to write templates. Reference~\cite{sstikettle}, also identifies two main SSTI scenarios: unintentional and intentional. The unintentional case is a web application that incorrectly embeds user inputs inside templates, while the intentional case is when the web application wants to allow users to interact with a template engine.

\section{Template Engines And Server-Side Template Injection}
\label{sec:teAndSSTI}

This section explains how template engines work, how SSTI arises, and why RCE is a common consequence. To this end, we will show practical code examples using \texttt{Jinja2}~\cite{jinja2docs} as a template engine, since it is one of the most popular. The section will be divided into five parts. In the first part (Section~\ref{sec:teAndSSTI:overview}), we comprehensively overview how template engines work. In the second part (Section~\ref{sec:teAndSSTI:whyRCE}), we explore the possible reasons why RCE happens in template engines. In the third part (Section~\ref{sec:teAndSSTI:rceExample}), we show an example of how SSTI can be exploited to achieve RCE. In the fourth part (Section~\ref{sec:teAndSSTI:SSTItypes}), we present the potential types of SSTI that can arise in web applications. Finally, in the fifth part (Section~\ref{sec:teAndSSTI:ThreatModel}), we provide a threat model showcasing the exploitation of an SSTI vulnerability.

\subsection{Overview of Template Engines}
\label{sec:teAndSSTI:overview}
To provide a general idea of how this technology works, we can structure a template engine into three components, as shown in Figure~\ref{fig:templateComponents}: \emph{(i)} A \emph{data source}, which is usually a database that can be used to retrieve information from a user's request; \emph{(ii)} The \emph{template engine} itself, which is the core that is responsible for taking the data extracted from a database and displaying it to the user on an HTML page (\texttt{index.html}). \emph{(iii)} The \emph{template page} (\texttt{index.tpl}), which is a custom HTML file containing \emph{symbols} the engine can interpret for specific functions. 

Template engines also allow programmers to include files, generate loops, use conditional statements (if-else), and call functions to transform or change the data before presenting it.

\begin{figure}[H]
\begin{tikzpicture}[node distance=2.6cm and 1.6cm]

\node (dataBox) [data] {user1 = 'John', user2 = 'Bob'};
\node (templateBox) [template, below=of dataBox] {\texttt{<p>\{\{x\}\} and \{\{y\}\}</p>}};
\node (engineBox) [engine, right=of $(dataBox)!0.5!(templateBox)$] {x=user1\\ y=user2};
\node (outputBox) [output, right=of engineBox] {\texttt{<p>John and Bob</p>}};

\node [above=2pt of dataBox] {Data};
\node [below=2pt of templateBox] {Template};
\node [above=2pt of engineBox] {Template Engine};
\node [above=2pt of outputBox] {Rendered HTML Output};

\draw [arrow] (dataBox) -- (engineBox);
\draw [arrow] (templateBox) -- (engineBox);
\draw [arrow] (engineBox) -- (outputBox);

\end{tikzpicture}
\caption{Components of an application that uses a template engine.}
\label{fig:templateComponents}
\end{figure}

To better understand template engines, we provide a simple example of how they work in practice. Consider a web application written in Python that uses \texttt{Jinja2} as a template engine, a popular engine integrated with the web framework Flask~\cite{flaskdocs}. Suppose a developer wants to embed a string sent by a user via a GET request in an HTML page. The correct way to use the template engine would be the following:
\begin{lstlisting}[language=Python, caption=Flask Jinja2 safe code example.,label={sec:overview:lst:backgroundJinja2SafeCode}]
user_input = request.form['username']
template = "<h1>Welcome, {{ user }}!</h1>"
render_template_string(template, user=user_input)
\end{lstlisting}

In the first line, we collect the user input through a form. We need to treat this input safely to avoid any injections. To pass the user input to the template, we use the argument \texttt{user=user\_input}, binding the \texttt{user} variable of our template to the Python variable \texttt{user\_input}. \texttt{Jinja2} will then execute the template and replace the \texttt{\{\{~user~\}\}} with the content of the \texttt{user\_input} variable. If \texttt{user\_input} contains John, the generated HTML page would contain a heading with the string \texttt{Welcome, John!}.

Conversely, the following code is vulnerable to SSTI. 
\begin{lstlisting}[language=Python, caption=Flask Jinja2 vulnerable code example.,label={sec:overview:lst:backgroundJinja2VulnerableCode}]
user_input = request.form['username']
template = "<h1>Welcome, %s!</h1>" % user_input
render_template_string(template)
\end{lstlisting}

The main difference between this code and the previous one is that, in this case, the user input is placed inside the template. This means the engine will interpret the user input as part of the template, not as separate data. The consequences of a vulnerable template are not immediately visible for normal inputs. For example, if \texttt{user\_input} was again containing \texttt{John}, we would again see the heading \texttt{Welcome, John!} rendered in the HTML. However, if instead of John our input was \texttt{\{\{~7*7~\}\}}, we would obtain a heading containing \texttt{Welcome, 49!}. This happens because the \texttt{Jinja2} template engine finds the curly brackets and executes the statement inside them.

The concept is similar to other injection-based vulnerabilities, such as SQL injection, which arises when we concatenate a user input to a query. SQL injections are mitigated by using prepared statements to pass unsanitized user inputs. To prevent SSTI, we should use the proper function arguments to safely pass the user-supplied data to the engine, as shown in the previous code (Listing~\ref{sec:overview:lst:backgroundJinja2SafeCode}). However, SSTI is potentially more severe in various cases than SQL injection, as it often leads to RCE, while SQL injection rarely does.

\subsection{Why is RCE Often Allowed?}
\label{sec:teAndSSTI:whyRCE}
Despite their apparently simple goal, template engines need to perform non-trivial operations in a secure way. Although Section~\ref{sec:teAndSSTI:overview}, showed a basic example of how template engines can be used, there are much more complex scenarios. An example can be provided by simply considering the possibility of using objects, attributes, and functions. Suppose we have a class called User in our web application. This class has various attributes, including username, first name, last name, and date of birth. If we want to create a template that contains this data using what we saw previously, we would have to write the following code:
\begin{lstlisting}[language=Python, caption=Flask Jinja2 object usage example (server-side).,label=sec:whyRCE:lst:noAttributesExample]
user = User('John98', 'John', 'Doe', '19/04/1998')
template = "<h1>Welcome, {{username}}!</h1><p>{{firstname}} {{lastname}} {{dateofbirth}}</p>"
render_template_string(template, username=user.username, firstname=user.firstname, lastname=user.lastname, dateofbirth=user.dateofbirth)
\end{lstlisting}

We can see that we can still easily achieve our goal, but the server code becomes longer. Now, let us show how cleaner the code would be by being able to access an object attribute directly inside the template.

\begin{lstlisting}[language=Python, caption=Flask Jinja2 object Usage example (template-side).,label=sec:whyRCE:lst:AttributesExample]
user = User('John98', 'John', 'Doe', '19/04/1998')
template = "<h1>Welcome, {{user.username}}!</h1><p>{{user.firstname}} {{user.lastname}} {{user.dateofbirth}}</p>"
render_template_string(template, user=user)
\end{lstlisting}

The \texttt{render\_template\_string} call is much less heavy, and considering that templates are usually contained inside separate files, we would have a very simple server-side code to render our template. Suppose that, in our web application, we want to check if a user has a premium account and display the page differently based on this information. In particular, to check if a user is a premium, we want to call the function \texttt{isPremium} on the \texttt{user} object. If template engines did not allow us to call functions or use conditional statements, we would have written the following server-side code:

\begin{lstlisting}[language=Python, caption=Flask Jinja2 conditions and functions usage example (server-side).,label=sec:whyRCE:lst:noFunctionsExample]
user = User('John98', 'John', 'Doe', '19/04/1998')
if user.isPremium():
    template = "<h1>Congratulations {{user.username}} you are now a premium member!</h1>"
else:
    template = "<h1>Welcome, {{user.username}}</h1>"
render_template_string(template, user=user)
\end{lstlisting}

As we can see in the above code, we would have to put this logic on the server-side code, making it more complex. Fortunately, template engines allow both functions and conditional statements, thus performing this task easily with the following code:

\begin{lstlisting}[language=Python, caption=Flask Jinja2 conditions and functions usage example (template-side).,label=sec:whyRCE:lst:functionExample]
user = User('John98', 'John', 'Doe', '19/04/1998')
template = "<h1>{% if user.isPremium() %}
                Congratulations {{user.username}} you are now a premium member!
            {% else %}
                Welcome, {{user.username}}
            {% endif %}</h1>"
render_template_string(template, user=user)
\end{lstlisting}

Again, since templates are usually defined in separate files, our server-side code would be much cleaner. Now, let us consider the price we pay in security for these valuable functionalities. By granting access to an object's attributes, we inadvertently expose special internal attributes that the object uses. In Python, for instance, introspective attributes are employed, enabling the traversal of objects and access to potentially risky modules, classes, and functions. Additionally, permitting the invocation of functions opens the door to system operations that could execute arbitrary commands, such as the \texttt{system} function.
If we developed our own template engine, we would face a critical decision: to allow unrestricted access to all attributes and functions or impose restrictions. 

In the next sections, we will show that the prevailing approach for most template engines is to grant templates access to all attributes and functions, often with the expectation that web developers will exercise caution to prevent Server-Side Template Injection (SSTI). However, this approach raises several concerns.
Firstly, security should not be deferred to others in the development process; it must be an integral part of every stage. Second, modern web applications sometimes necessitate user access to the template engine. A prominent example is a Content Management System (CMS) framework and platforms like Github Pages or bulk email services that allow clients to utilize template engines without undue restrictions. In such cases, avoiding RCE is paramount.
Comparing SQL injection and SSTI reveals a significant distinction. In the case of SQL injection, it's generally agreed that the responsibility lies with the web application, not the Database Management System (DBMS). Conversely, with SSTI, we find that some template engines facilitate RCE while others do not. The problem remains that RCE via SSTI is a critical concern that persists in web applications. Therefore, web developers must be aware of these risks and know how to properly employ and configure template engines to prevent RCE effectively.

\subsection{RCE Example}
\label{sec:teAndSSTI:rceExample}
The code in Section~\ref{sec:teAndSSTI:overview} showed an unsafe method for web developers to use template engines because malicious users could inject template syntax to send commands that the server would execute. Since Python has attributes and methods that can be used to access the global scope of the application and execute arbitrary functions from imported modules, attackers can easily develop an \emph{introspective payload}. This kind of payload begins by exploiting an \emph{object}. In the following example, we used the Flask \emph{config} object, but we can use any object.
\begin{lstlisting}[language=Python,caption=Introspective payload that allows RCE in Jinja2 exploiting the config object.,label={sec:rceExample:lst:resultsJinja2PayloadConfig}, morekeywords=config]
{{config.__class__.__init__.__globals__['os'].popen('ls').read()}}
\end{lstlisting}
The payload above is one of many that can be used to exploit SSTI in \texttt{Jinja2}. In this case, the payload starts with accessing \texttt{config}, a \texttt{Flask} global object that contains sensitive server information. We can use this object to traverse through its attributes by accessing \texttt{\_\_class\_\_} first and the method \texttt{\_\_init\_\_} to reach the object \texttt{\_\_globals\_\_}, which is a dictionary that contains every object inside the scope of the web server. From here, we can access the \texttt{os} module, which allows the execution of arbitrary commands with the function \texttt{popen}. The payload will execute the \texttt{ls} command in this case. We use the final call to the \texttt{read} function to see the command's output on the web page. 
\begin{lstlisting}[language=Python,caption=Generic Introspective payload that allows RCE in Jinja2. N is the index where the class Popen resides in the list returned by the \_\_subclassess\_\_ function call.,label={sec:rceExample:lst:resultsJinja2PayloadString}, morekeywords=N]
{{''.__class__.mro()[1].__subclasses__()[N]('ls',shell=True,stdout=-1)}}
\end{lstlisting}
This payload distinguishes itself from the one in Listing~\ref{sec:rceExample:lst:resultsJinja2PayloadConfig} due to its initial use of an empty string. This distinction makes the payload less reliant on specific objects, enhancing flexibility. However, it remains essential to utilize the \texttt{\_\_class\_\_} attribute, as it allows access to the \texttt{mro} method, which, in turn, provides a hierarchy of the parent and child classes of \texttt{String}. This is useful because our aim is to reach the class \texttt{Object}. Since all classes inherit from \texttt{Object}, this can be done from any class.
The class \texttt{object}, in this case, is positioned within this hierarchy at index one (\texttt{String} is at zero). Once we have a reference to the \texttt{Object} class,  we can access the \texttt{\_\_subclasses\_\_()} method. This method furnishes a list of all subclasses of an object, effectively exposing all the classes accessible within the web server's context. This is already something undesirable because it allows an attacker to access dangerous classes.

Now, the attacker can escalate even further, locating the \texttt{Popen} class among the hundreds of values returned from \texttt{\_\_subclasses\_\_}.
It is worth noting that determining the \emph{N} index necessitates careful consideration, as it depends on the location of the \texttt{Popen} class within the web application's context. This index varies according to the modules, classes, and functions imported and utilized within the web application. Attackers can either bruteforce this index until they find the correct class or print the result of \texttt{\_\_subclasses\_\_()} to calculate the correct index.

Ultimately, this payload creates a \texttt{subprocess} object that executes the \texttt{ls} command. Notably, this approach can be adapted to execute arbitrary code or commands, offering a broad range of possibilities for attackers to take control of the server.

Now that we have seen the details on how SSTI can be escalated to RCE, we can explore the different ways in which SSTI arises.

\subsection{Types of SSTI}
\label{sec:teAndSSTI:SSTItypes}
As with other web-based vulnerabilities like Cross-Site Scripting (XSS), SSTI can also be triggered in different ways. The three main identified types encompass distinct scenarios in which SSTI arises.
\begin{itemize}
    \item \textbf{Non-persistent}. The SSTI payload is sent by the attacker and embedded inside the template. The response is then rendered, and the payload is executed.
    \item \textbf{Persistent}. The SSTI payload is sent by the attacker and stored on the web application servers. The payload is then retrieved and embedded inside the template. The attack is executed when the attacker triggers the loading of the payload inside a template.
    \item \textbf{Blind}. If the payload execution result is not shown on the rendered page or anywhere else, we are in a blind scenario. The attacker can exfiltrate data from the server using an out-of-band communication. An example of blind SSTI can be when a template is parsed but not rendered on the webpage.
\end{itemize}

Considering the above types of SSTI, we can have four combinations of different scenarios: (i) Non-blind non-persistent SSTI, (ii) Non-blind persistent SSTI, (iii) blind non-persistent SSTI, and finally (iiii) blind persistent SSTI.

\subsection{Threat Model}
\label{sec:teAndSSTI:ThreatModel}
The exploitation of an SSTI vulnerability begins with the attacker injecting custom payloads that are interpreted by the application as part of the template. These malicious payloads can be delivered through various parts of the request, although the GET and POST parameters of HTTP requests are the primary delivery methods. The payload may be extracted from these attributes and reflected within a template declaration, indicating to the attacker that the payload was executed successfully. At this stage, the attacker may be able to trigger an XSS attack by injecting an HTML string into the SSTI payload—providing an initial, tangible consequence that can be exploited.

An attacker can also achieve remote code execution (RCE) by exploiting SSTI, depending on whether the template engine supports arbitrary code execution. As shown in Table~\ref{tab:popularity}, most template engines allow RCE, which significantly increases both the likelihood and the impact of SSTI exploitation.

It is also possible for an application to persistently store an SSTI payload, which is later retrieved and executed when a specific part of the website is accessed. In such cases, an attacker may exploit features that allow users to create example templates without being aware that RCE is possible within them. This attacker, who may be a legitimate user of the functionality but is not supposed to have server-level access, can exploit the feature to gain unauthorized control of the server.

An expert attacker may also discover a blind SSTI vulnerability by executing tailored payloads that trigger differences in the server response times. After discovering and confirming the SSTI, the attacker can use sophisticated payloads to exfiltrate the data without having a visual response from the server.

\section{Remote Code Execution in Template Engines}
\label{sec:RCEinTE}
This section explores the different categories of RCE vulnerabilities in template engines and the corresponding protective measures. To this aim, we will start by categorizing the differences in how developers address RCE in their engines by describing the various ways in which it arises. This part of the paper is based less on related academic papers (since no other paper analyzes template engines in this way) and more on the results of experiments that we carried out on a set of 34 selected template engines.

\subsection{RCE Types in Template Engines}
\label{sec:RCEinTE:RCEtypes}

The RCE example in Section~\ref{sec:teAndSSTI:rceExample} is just one of the many ways in which RCE can be achieved in template engines. After analyzing 34 template engines, we have found four main categories of RCE exploits. We will now present this categorization of exploits to understand how template engines allow RCE in different ways. 
\begin{itemize}
    \item \textbf{Direct code execution}. The execution of any code that is embedded inside the template engine's default delimiters. This practice makes the template engine extremely dangerous in case of a template injection, as an attacker can easily and immediately execute arbitrary code.
    \item \textbf{Tags or functions for code execution}. The presence of \emph{specific delimiters, functions, or instructions} in the template engines that can be used to execute arbitrary code. This is a deliberate feature that engine creators provide because they believe web developers might want to use them to execute specific pieces of code inside the engine (e.g., define helper functions). Even if there is nothing wrong with providing such features, web developers should be aware of this kind of danger if SSTI arises or if they want to allow users to use the engine inside their web application.
    \item \textbf{Introspective}. An \emph{introspective} exploit depends on the programming language of the template engine. Each language has different functions and attributes that can be exploited to traverse objects and achieve arbitrary code execution. This exploit is rather common as template engines typically allow access to attributes and functions on the objects passed to them. Therefore, using introspective features of object-oriented languages is the easiest way to attack this mechanism.
    
    \item \textbf{Bugs or vulnerabilities}. Bugs or vulnerabilities can arise in the template engine and be turned into a way to achieve RCE. This category encompasses unintended behaviors of template engines that can be exploited to achieve RCE in engines that would otherwise be secure from it.
\end{itemize}

\subsection{Preventing RCE}
\label{sec:RCEinTE:preventRCE}
Some template engines also enforce \emph{security features} that allow for better resistance against specific exploits. These features are usually implemented by engine developers to block or reduce the risk of RCE. We use the term resistance because these security features are designed to provide higher protection against RCE. However, their effectiveness depends on how well they are implemented and if they are enforced by default, or if they have an opt-in switch. In the following, we show the various security features that can be provided in template engines.
\begin{itemize}
\item \textbf{Sandbox}. The easiest way to patch or limit template engines' arbitrary code execution capabilities is to prevent certain actions or block specific functions. Hence, sandboxes are used by the majority of the template engines that want to provide some basic level of security. We can define this feature as the existence of a \emph{blacklist} of words in the syntax of the template engine. Whenever a word inside the blacklist is found, the execution of the template is blocked. The problem with this kind of security feature is that it does not provide a robust way to protect against arbitrary code execution. Sandboxes are escaped in many similar contexts, so it should be known to web developers that this kind of risk can be avoided by only keeping the engine updated. Besides, many template engines only put sandboxes in place after discovering an RCE exploit, making previous versions vulnerable. This is another fact that should be known to developers. If a template engine provides a sandbox, the latest version of the engine should always be on, or there is a risk that the sandbox is not updated.

\item \textbf{No function calls}. All the introspective payloads found for the various template engines require a function call at a certain point, either to instantiate an arbitrary object or to call an introspective function. Blocking any function call can effectively avoid code execution but it might also greatly impact the template engine functionalities.

\item \textbf{Limited code execution}. Some template engines greatly limit what the user can or cannot do inside the delimiters. We could also say that sandboxes are meant to limit code execution. However, the main difference between these two categories is that sandboxes limit specific functions or attributes to systematically block specific payloads that can be exploited to achieve RCE. In this case, we refer to a more general limit that can be either imposed by blocking most functions or attribute access (with a whitelist, for example) or by giving only a set of custom directives that the developer can use in the template.

\item \textbf{No RCE vulnerability}. Some template engines might not have any specific security feature, but no arbitrary code execution was found. An example where this can happen is when the language of the template engine cannot have introspective payloads.
\end{itemize}

\section{Literature Review}
\label{sec:literature}

To identify which works related to SSTI have been published, we conducted keyword searches for terms such as "server-side template injection," "SSTI," or "template injection" across various search engines for our research, our investigation revealed that Google Scholar proved to be the most comprehensive source, yielding the majority of relevant papers. Other search engines, namely IEEE, ACM, and SCOPUS, either failed to find any results related to SSTI or duplicated the same works we had already identified on Google Scholar. Our analysis revealed only four papers related to SSTI.

In this section, we delve into three of these papers, while the fourth~\cite{zapscanner} is reviewed in Section~\ref{sec:detectionTools} since it is a paper that presents an SSTI detection tool. The three papers that we review in this section are the following. (i) Kettle's paper regarding SSTI discovery and exploitation~\cite{sstikettle}; (ii) Wang et. al.~\cite{instruction_randomization} paper that proposes to protect against SSTI with instruction set randomization; (iii) Zhao et. al.~\cite{zhaoremote} paper that shows how sandboxed PHP template engines could be exploited to achieve RCE. 

To the best of our knowledge, the concept of SSTI was first brought to light in 2015 by James Kettle~\cite{sstikettle}. Our research uncovered no references to this vulnerability in the academic literature before that year.

His research showed for the first time the critical consequences associated with template engines and how attackers can leverage them to achieve Remote Code Execution (RCE). James Kettle's work on SSTI was characterized by its analysis of the vulnerability, the practical exploitation techniques, and the far-reaching implications of SSTI in the context of modern web applications. 
Furthermore, Kettle analyzed five template engines (FreeMarker, Velocity, Smarty, Twig, and Jade), showing how they allowed RCE and how sandbox protections could be bypassed. Finally, his work showed case studies of prominent frameworks and CMS applications that are vulnerable to SSTI, demonstrating the possible prevalence of this vulnerability. Table~\ref{tab:literature} summarizes the literature related to SSTI from 2015 until now.

\begin{table}[H]
\begin{tabular}{@{}cccc@{}}
\toprule
\textbf{Ref.}                  & \textbf{Tool} & \textbf{Year} & \textbf{Objective}                                        \\ \midrule
\cite{sstikettle}                & Burp plugin   & 2015          & Detection and exploitation of SSTI                        \\
\cite{zapscanner}                & ZAP plugin    & 2018          & A tool for SSTI detection that uses polyglot payloads     \\
\cite{instruction_randomization} & -             & 2021          & Defending against SSTI with instruction set randomization \\
\cite{zhaoremote}                & TEFuzz        & 2023          & Fuzzing PHP template engines to escape sandboxes          \\ \bottomrule
\end{tabular}
\caption{Overview of SSTI-related literature}
\label{tab:literature}
\end{table}

\subsection{Kettle's Exploitation Methodology}
\label{sec:literature:exploitMethodology}
In Reference~\cite{sstikettle}, Kettle provides a step-by-step methodology to exploit SSTI in a black-box scenario. This methodology is also used by SSTI detection tools and is useful to understand the differences and similarities between template engines and how important it is, on the exploitation side, to know which engine a web application is running. The methodology is based on three main steps:
\begin{itemize}
    \item \textbf{Detect}. In the detection phase, the goal is to confirm the presence of SSTI. There are two general contexts in which SSTI can arise: plaintext and code. In the plaintext context (we saw an example in Listing~\ref{sec:overview:lst:backgroundJinja2VulnerableCode}), HTML can be directly input into templates, and it's often a source of XSS. To detect this, one can invoke the template engine with generic payloads (e.g., the typical \{\{7*7\}\}). 
    
    In the code context, user input is placed within template statements instead, making it less obvious to discover. An example is the following code:

    \begin{lstlisting}[language=Python,caption=Flask Jinja2 code context SSTI example.,label={sec:exploitMethodology:lst:SSTICodeContextExample}]
        user_input = request.form['username']
        template = "<h1>Welcome, {{%s}}!</h1>" % user_input
        render_template_string(template)
    \end{lstlisting}
    This scenario can be discovered by trying to close the statement and injecting HTML tags afterward (e.g. \}\}<p>SSTI</p>).
    \item \textbf{Identify}. Once SSTI is detected, the next step is identifying the template engine in use. Sometimes, sending invalid syntax to cause error messages can reveal the engine. Otherwise, it is necessary to try different payloads and observe whether they are executed.
    \item \textbf{Exploit}. After finding the template injection and identifying the template engine, attackers usually need to find a way to escalate to RCE. The first step can be to read the template engine documentation since it can provide insightful information on methods, variables, and special functionalities. If a way to escalate has not been found at this point, then the engine environment can be explored to see if it leaks any sensitive information that can be used to further compromise the system. Additionally, templates might expose objects that can be exploited by an attacker to perform malicious actions.
\end{itemize}
Despite the utility and practicality of this methodology, it mainly provides an offensive side of the SSTI vulnerability. Furthermore, it lacks generalization for the widespread variety of template engines that we have nowadays. 
Future research could work on both providing a more general method to detect and protect against SSTI vulnerabilities and exploring automated approaches for assessing RCE in template engines.

\subsection{Protecting Against SSTI With Instruction Set Randomization}
\label{sec:literature:ISR}
Wang et al.~\cite{instruction_randomization} proposed a new method for defending against Server-Side Template Injection (SSTI), which circumvents the need for specialized tools or scanners. This method is already used for other vulnerabilities~\cite{kc2003CCS, boyd2008TDSC, sinha2017HOST} and is known as instruction set randomization~\cite{instruction_randomization}. This technique introduces an element of randomness into the template code by incorporating a random key. Notably, template engines rely on specific delimiters and instructions to parse templates. For instance, when the engine encounters the string \texttt{\{\{} (commonly used as the default delimiter in template engines like \texttt{Jinja2} and others), it identifies this as the beginning of a variable block to be executed. Some template engines provide environment functions or attributes that allow the customization of these delimiters. In cases where these options are not readily provided, the template engine's code can usually be modified to recognize different characters as the instruction delimiter (e.g., instead of using \texttt{\{\{ \}\}} as delimiters, alternative symbols like \texttt{[[ ]]} could be employed). 

Instruction set randomization is based on this functionality and works by generating a random pair of delimiters (e.g., \texttt{\{\{randomString \}\}}) that is unknown to potential attackers. When a web application is vulnerable to SSTI, an attacker attempting to inject template code would fail to execute it, as the engine would not recognize the default delimiters. It is important to note that the effectiveness of this protection relies on the secrecy of the selected or generated random string. Therefore, the string should not be leaked in any way on the application, and it needs to be sufficiently long and complex to resist brute-force attacks.

This SSTI defense approach is especially effective in scenarios where SSTI is unintentional. In cases where a web application intentionally exposes the template engine to users, this protection becomes impractical. Moreover, applying this technique in cases of unintended SSTI can impose a substantial burden on developers, as they would need to manually modify each template by adding the random string before every instruction (a process that could potentially be automated). Nevertheless, automation would require the creation of scripts to parse template files and implement the random string addition. Altering delimiters may not always be straightforward: as we said, many templates offer customization options, but others may necessitate changes to the template engine's core code. This practice introduces several security risks, including the need to re-modify the code with every engine update and the potential for introducing new vulnerabilities during code modifications.

In conclusion, the instruction set randomization approach offers a unique perspective on defending against SSTI but may not necessarily surpass other strategies or tools. In the future, this research path can be further explored to find automated ways to implement instruction set randomization that are reliable and easy to implement for developers.

\subsection{PHP Template Engines Sandbox Escaping}
\label{sec:literature:escapingTheSandbox}
In section~\ref{sec:RCEinTE:preventRCE}, we saw that template engine developers have introduced sandbox modes to restrict the capabilities of template tags, preventing attackers from achieving RCE. However, a paper by Zhao Y. et al.~\cite{zhaoremote} recently presented a fuzzer that aims to automatically detect and exploit sandbox escape bugs in template engines. In fact, it exploits a previously overlooked bug called \emph{template escape}, which allows attackers to bypass the sandbox and gain RCE. This bug occurs when attacker-controlled inputs in the template code escape the template's intended semantics during translation to PHP code. Reference~\cite{zhaoremote}, uses CVE-2021-26120 as an example to illustrate how template escapes bugs work. By carefully crafting a template function name, attackers can inject PHP code into the translated PHP file, thereby gaining execution. The paper suggests that template escape bugs may not be as rare as previously thought and calls for a study of their prevalence and severity in every template engine.

Despite focusing only on a specific set of PHP engines, this work shows an increasing interest in SSTI research. The results obtained by this paper also underline how SSTI still affects many CMSs and how sandboxing mechanisms cannot be trusted. TEFuzz is a tool that works with Tplmap and aims to detect sandbox escapes in PHP template engines. The results show that applying TEFuzz to 7 PHP engines, 55 bugs can be exploited to gain RCE on sandbox-protected template engines. 

Also, Reference~\cite{sstikettle}, discussed possible sandbox bypasses that existed in various template engines. Despite the majority of the sandbox bypasses shown being in template engines written in Java or PHP, it is possible that other engines in other languages can be vulnerable. The possibility that sandboxes are evaded (and therefore become useless in defending against RCE) is a worrying sign that this kind of protection might not be ideal. This calls for future research to uncover possible sandbox escapes in other engines and propose alternative ways to protect against RCE.

\subsection{Open Research Gaps}
\label{sec:literature:researchGaps}

As highlighted in previous sections, research on SSTI remains limited. Nevertheless, the works analyzed in this section have opened several promising topics for further investigation.

\textbf{SSTI Detection and Prevention}.
Despite the existence of some detection tools, which will be further discussed in the next section, and the work by Silva~\cite{zapscanner}, SSTI detection largely relies on black-box testing through payload injection on target websites. Currently, there are no white-box or gray-box detection tools available, nor have any research papers addressed this approach. On the prevention side, while the work by Wang et al.~\cite{instruction_randomization} on instruction set randomization is notable, a practical or comprehensive implementation strategy is still lacking.

\textbf{RCE Detection and Prevention}.
Kettle~\cite{sstikettle} and Zhao~\cite{zhaoremote} have both shown that sandbox mechanisms in template engines are unreliable and can be bypassed. However, automatic detection of this issue has only been addressed for PHP-based engines, leaving engines for other programming languages largely unexplored. Furthermore, the prevention of RCE remains a critical yet neglected area within SSTI research, with no published work offering a detailed analysis or solutions. There is a clear need for the development of methods to automatically block, detect, or even patch RCE vulnerabilities in template engines.

\section{SSTI Detection Tools}
\label{sec:detectionTools}
 
This section will summarize how the three most popular tools to detect SSTI work and their main differences. This comparison can be useful to identify strengths, pitfalls, and improvements that future tools might implement. 
We can evaluate their popularity to make a selection among these tools. Burp stands out as a widely recognized commercial vulnerability scanner, while ZAP is the leading open-source alternative. Furthermore, Tplmap has gained popularity, evident from its 3.5k GitHub stars and its status as an official Burp extension. It's worth noting that, during our search for "SSTI" or "template injection" tools on GitHub, we observed that many others are either forks or draw inspiration from Tplmap. Among the three tools we analyzed, the first one to be released, immediately after the discovery of SSTI by J. Kettle, was an extension to detect SSTI automatically for \texttt{Burp Suite}. Since \texttt{Burp}~\cite{burpsuite} is one of the most popular web application security testing software, the vulnerability started to get some attention. After that, two other tools were created to detect and exploit SSTI.  Table~\ref{tab:tools} shows a general comparison between the four tools described in detail in the following. Notably, analyzing the Burp scanner's details is not possible since it is not open-source, so our analysis will focus on Tplmap, ZAP-Esup, and SSTImap.

\begin{table}[]
\begin{tabular}{@{}cccccc@{}}
\toprule
\textbf{Name}         & \textbf{Year}            & \textbf{Open-source} & \textbf{Supported engines} & \textbf{Plugin} & \textbf{Crawler} \\ \midrule
\textbf{Burp scanner} & 2015                     & \faTimes             & -                            & Burp            & \faCheck         \\
\textbf{Tplmap}       & 2016                     & \faCheck             & 18                           & Burp            & \faTimes         \\
\textbf{ZAP-ESUP}     & 2018                     & \faCheck             & 18                            & ZAP             & \faCheck         \\
\textbf{SSTImap}      & 2022 & \faCheck             & 27                           & -               & \faTimes         \\ \bottomrule
\end{tabular}
\caption{SSTI detection tools comparison. The number of template engines supported by \texttt{Burp} is unknown.}
\label{tab:tools}
\end{table}

\subsection{Tplmap} 
\label{sec:detectionTools:tplmap}
\texttt{Tplmap}~\cite{tplmap} is a command-line tool designed for detecting and exploiting SSTI vulnerabilities in web applications. Once a user detects a potential SSTI, Tplmap can automate the exploitation process by injecting payloads into the application to assess whether the vulnerability can be leveraged for RCE. Tplmap supports various template engines commonly used in web development, such as \texttt{Jinja2}, \texttt{Smarty}, and \texttt{Twig}, and it can detect and exploit SSTI vulnerabilities specific to a limited set of engines. Users can also provide custom payloads to Tplmap that can tailor the exploitation attempts to specific needs or test against unique scenarios. The main limitation of this tool is the restricted amount of template engines supported. \texttt{Tplmap} supports several popular template engines but does not cover every template engine used across the web. This means that some applications using less common or custom template engines may not be effectively tested using this tool. At this moment, \texttt{Tplmap} supports 18 template engines. Secondly, it lacks crawling capabilities: \texttt{Tplmap} cannot automatically find vulnerable endpoints, they have to be specified by the user. Furthermore, its repository is no longer maintained, thus lacking support for recent template engines. This, along with the aforementioned limitations, calls for new tools that will be able to keep up with the evolving landscape of SSTI and the ever-rising presence of new template engines.

\subsection{ZAP-Esup}
\label{sec:detectionTools:zapEsup}
\texttt{ZAP-ESUP}~\cite{zapscanner} was published in 2018 by Diogo Silva as an extension for Zed Attack Proxy (ZAP)~\cite{zaproxy}. The main feature of this tool is that it uses polyglot payloads to overcome the limitations of \texttt{Tplmap} in terms of template engines supported and exploitation methodology. With a polyglot payload, in this case, the author means a string of symbols and letters that can help detect SSTI in multiple template engines.

The paper also discusses the development of an SSTI polyglot payload designed to cause errors in web applications. To determine the most effective method for triggering errors, the author tested four combinations of template tags: start tag (e.g., \texttt{\$\{}), ending tag (e.g., \texttt{\}}), start tag followed by an ending tag (e.g., \texttt{\$\{\}}), and start tag with a variable name and ending tag (e.g., \texttt{\$\{foo\}}). They found that the best way to cause errors was by sending a start tag, a nonexistent variable name, and an ending tag, resulting in unusual behavior in 16 out of 18 tested applications. To detect unusual behaviors, the tool registers how the application responds with normal inputs and then with the SSTI payload. In this case, the absence of the variable in the template engine context causes an error, resulting in a different server response.

However, Java-based applications handle errors better with respect to other programming languages, without causing differences in the response when variables do not exist. To address this, the author creates a specific polyglot for Java applications that contains code intentionally violating their syntax to trigger errors. Then, to reduce the payload size, they create a general polyglot by combining various template start and end tags around an existing payload. Finally, they backslash the general polyglot to prevent rendering and add \texttt{Twig} commentary tags (\texttt{\{\# \}}) to address issues with \texttt{Smarty}. The goal is to create an effective SSTI polyglot that can trigger errors in different web applications, thus detecting SSTI without specific engine knowledge.

\texttt{ZAP-ESUP} provides various improvements concerning \texttt{Tplmap} and handles some of the limitations we previously showed. Despite these improvements and the fact that it can detect SSTI without relying on specific engine payloads, it still does not consider possible new engines with syntaxes that are not supported by the current polyglot payloads. The state of the art also lacks unbiased tests to assess which of these two tools can detect SSTI more efficiently.

\subsection{SSTImap}
\label{sec:detectionTools:sstimap}
SSTImap is a tool based on Tplmap but with some differences in terms of functions and supported templates. It has the drawback of not being integrated as a Burp plugin, while Tplmap supports this integration. Similarly to Tplmap, it does not yet support a crawling integration, even though it appears to be a future plan. In terms of improvements, SSTImap provides support for 27 engines and contexts. Even though they are more than Tplmap it is still a small number, especially if we consider that they are not 27 different template engines but some of them are eval-like code injections or payloads for different versions of the same engine. The way in which it works is similar to Tplmap, it injects a set of payloads on the target URL and attempts to detect which template engine is being used.

\section{Template Engine analysis}
\label{sec:templateAnalysis}
We highlighted the importance of selecting a template engine and knowing the security risks related to each engine. This section provides a methodology to analyze template engines and assess their security. We selected 34 template engines to analyze based on their popularity, searching both on GitHub (using stars as a measure of popularity) and other search engines (exploring top results for queries like "template engines in" plus the programming language of choice). 

\subsection{Programming Languages}
\label{sec:templateAnalysis:programmingLanguages}
The automatic analysis of template engines for RCE paths presents a challenging issue due to the wide variety of template engines available. This diversity directly results from the numerous programming languages used for web application development and frameworks. Template engines operate as a series of Application Programming Interfaces (APIs) that can only be used by the same programming language in which the template engine is written. Consequently, a template engine written in \texttt{Python} can only be used by \texttt{Python} web applications. 

\begin{figure}[h]
\includegraphics[width=\textwidth]{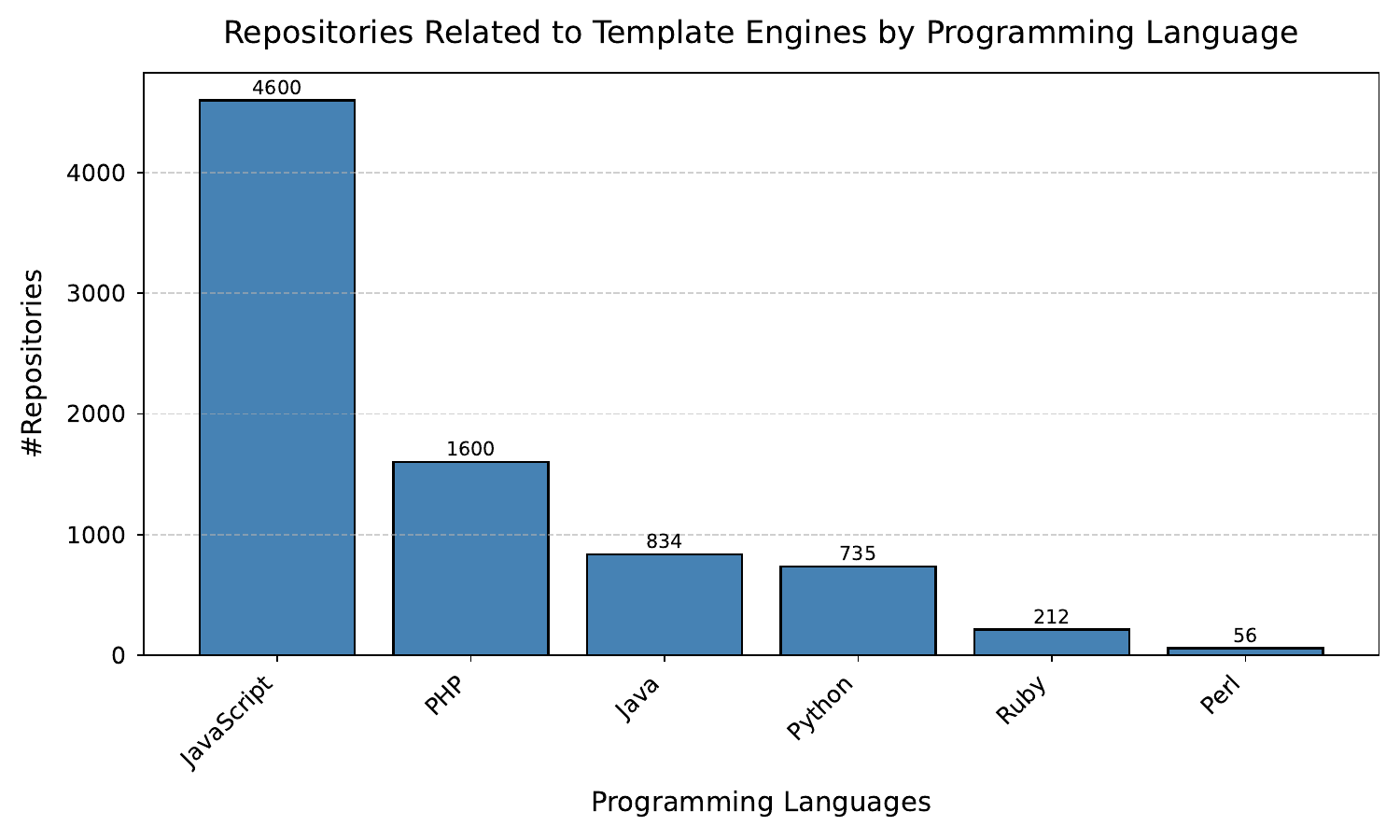}
\caption{Estimated number of template engines related repositories for each programming language by searching \emph{template engine} on GitHub.}
\label{fig:reposByLanguage}
\end{figure}

The complexity in automatically detecting RCE also arises because the payloads are linked to the programming language. An 
 introspective exploit that works for \texttt{Python} does not work for \texttt{Java} and vice versa. This variety poses many challenges in defending against RCE and SSTI, one being the difficulty in assessing if a template engine allows RCE without depending on the engine language.
 To understand how many different programming languages are used for developing template engines, Figure~\ref{fig:reposByLanguage} provides the results obtained when searching for \emph{template engine} on GitHub and reporting how many repositories had code in a certain language. We only considered some of the programming languages present in our analysis. The results show that six popular languages covered about $60$\% of the total repositories.

\subsection{Analysis Results}
\label{sec:templateAnalysis:analysisResults}
\begin{figure}[h]
\includegraphics[width=\textwidth]{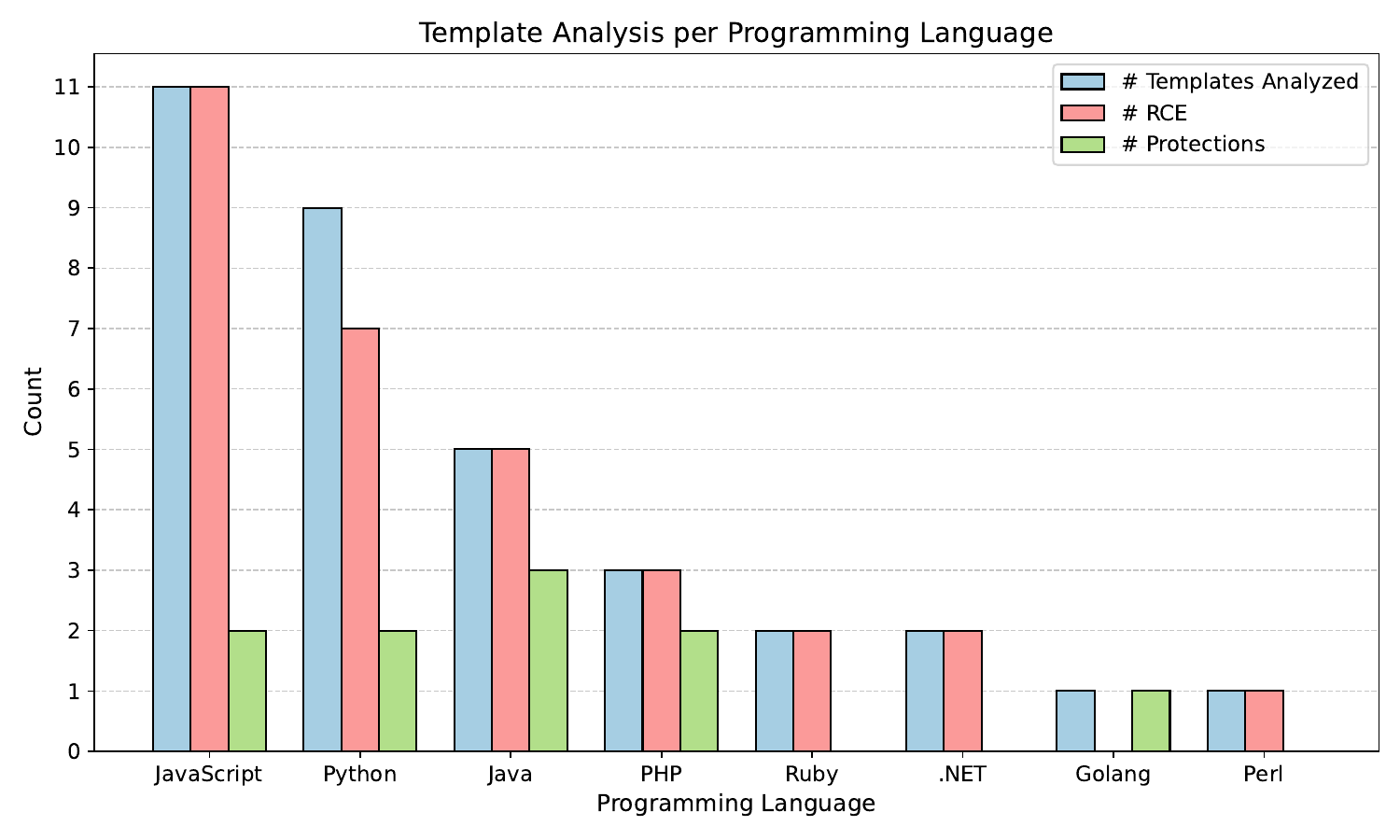}
\caption{Summary of how many Template Engines have or had RCE and how many implement any kind of protection to prevent it.}
\label{fig:templateOverview}
\end{figure}

To provide an overview of the current state of template engines regarding RCE and protections against its threat, we propose an analysis of $34$ template engines in eight different programming languages. Table~\ref{tab:templates-analyzed} shows a general overview of the results, while Figure~\ref{fig:templateOverview} shows some statistics extracted from them. The analysis was carried out with a methodology that involves three main steps:
\begin{enumerate}
    \item \textbf{Developing a minimal SSTI-vulnerable application.} The first step in analyzing an unknown template engine is to create an environment where we can test the template in an SSTI scenario. More specifically, the main objective of this phase is to create a \emph{vulnerable} piece of code by progressively modifying existing snippets to trigger SSTI. Such snippets can often be found in the documentation of the template engine and can be modified to produce a vulnerable scenario. With this code, we can proceed to test possible RCE payloads. 
    \item \textbf{Testing exploits and security features.} We can test possible exploits after creating a working environment with a running SSTI-vulnerable code. To this end, we devised a \emph{multi-step} procedure that can reveal whether the template engine is exploitable through RCE. This procedure is organized into \emph{steps} of progressive complexity, where we design four exploitation methodologies, which we detail in Section~\ref{sec:RCEinTE:RCEtypes}. If we find a working exploit from one of the previous steps, we can consider the target template exploitable with RCE. However, if we cannot find any working exploit, it does not mean that the template engine is secure against RCE. In addition to that, we evaluate possible \emph{security features} that characterize the target template according to a taxonomy that we describe in Section~\ref{sec:RCEinTE:preventRCE}.
    \item \textbf{Collecting the results.} In the final step, we collect the previous analysis's results. This part is essential to build a standard set of information that can be consulted and updated by researchers or programmers. In the future, template engines will be updated, and there is the possibility that they will introduce security measures that might change their ability to execute arbitrary code. Sometimes, they might even introduce bugs or updates that allow code execution. The attained results are organized in a report that contains the following information: 
    \emph{(i)} \emph{Name and language} of the template engine; \emph{(ii)} \emph{delimiters} used by the engine; \emph{(iii)} type of \emph{working exploit}; \emph{(iiii)} the potential \emph{security features} employed by the template.
The report also briefly describes the engine, examples of secure and vulnerable codes, and a more detailed analysis of the payload that allows RCE and the security features for each template engine.
\end{enumerate}

\begin{table*}
\begin{tabular}{clccccc} 
\toprule
\textbf{Language}            & \multicolumn{1}{c}{\textbf{Name}}                      & \textbf{Delimiters}                                                 & \textbf{A. A.}                               & \textbf{E.F.}                                & \textbf{Exploit Kind}                                        & \textbf{Security Features}                             \\ 
\midrule
\multirow{9}{*}{Python}      & Django                                                 & \{\{ \}\}                                                           & \faCheck                                     & \faTimes                                     & -                                                            & LCE                                                    \\
                             & {\cellcolor[rgb]{0.949,0.949,0.949}}Tornado            & {\cellcolor[rgb]{0.949,0.949,0.949}}\{\{ \}\} and \{\% \%\}         & {\cellcolor[rgb]{0.949,0.949,0.949}}\faCheck & {\cellcolor[rgb]{0.949,0.949,0.949}}\faCheck & {\cellcolor[rgb]{0.949,0.949,0.949}}Tag code execution       & {\cellcolor[rgb]{0.949,0.949,0.949}}-                  \\
                             & Jinja2                                                 & \{\{ \}\}                                                           & \faCheck                                     & \faCheck                                     & Introspection                                                & -                                                      \\
                             & {\cellcolor[rgb]{0.949,0.949,0.949}}\textbf{web2py}    & {\cellcolor[rgb]{0.949,0.949,0.949}}\{\{= \}\}                      & {\cellcolor[rgb]{0.949,0.949,0.949}}\faTimes & {\cellcolor[rgb]{0.949,0.949,0.949}}\faCheck & {\cellcolor[rgb]{0.949,0.949,0.949}}Introspection            & {\cellcolor[rgb]{0.949,0.949,0.949}}-                  \\
                             & Mako                                                   & \textless\% \%\textgreater and \$                                   & \faCheck                                     & \faCheck                                     & Tag code execution                                           & -                                                      \\
                             & {\cellcolor[rgb]{0.949,0.949,0.949}}\textbf{Chameleon} & {\cellcolor[rgb]{0.949,0.949,0.949}}\$\{ \}                         & {\cellcolor[rgb]{0.949,0.949,0.949}}\faTimes & {\cellcolor[rgb]{0.949,0.949,0.949}}\faCheck & {\cellcolor[rgb]{0.949,0.949,0.949}}Introspection            & {\cellcolor[rgb]{0.949,0.949,0.949}}-                  \\
                             & \textbf{Cheetah3}                                      & \$ and \#                                                           & \faTimes                                     & \faCheck                                     & Tag code execution                                           & -                                                      \\
                             & {\cellcolor[rgb]{0.949,0.949,0.949}}\textbf{Genshi}    & {\cellcolor[rgb]{0.949,0.949,0.949}}\$ and \textless? ?\textgreater & {\cellcolor[rgb]{0.949,0.949,0.949}}\faTimes & {\cellcolor[rgb]{0.949,0.949,0.949}}\faCheck & {\cellcolor[rgb]{0.949,0.949,0.949}}Tag code execution       & {\cellcolor[rgb]{0.949,0.949,0.949}}-                  \\
                             & Pyratemp                                               & @! !@                                                               & \faTimes                                     & \faTimes                                     & -                                                            & Sandbox                                                \\ 
\midrule
\multirow{3}{*}{PHP}         & {\cellcolor[rgb]{0.949,0.949,0.949}}\textbf{Laravel}   & {\cellcolor[rgb]{0.949,0.949,0.949}}\{\{ \}\}                       & {\cellcolor[rgb]{0.949,0.949,0.949}}\faTimes & {\cellcolor[rgb]{0.949,0.949,0.949}}\faCheck & {\cellcolor[rgb]{0.949,0.949,0.949}}Tag code execution       & {\cellcolor[rgb]{0.949,0.949,0.949}}-                  \\
                             & Twig                                                   & \{\{ \}\}                                                           & \faCheck                                     & \faCheck                                     & Introspection                                                & Sandbox                                                \\
                             & {\cellcolor[rgb]{0.949,0.949,0.949}}Smarty             & {\cellcolor[rgb]{0.949,0.949,0.949}}\{ \}                           & {\cellcolor[rgb]{0.949,0.949,0.949}}\faCheck & {\cellcolor[rgb]{0.949,0.949,0.949}}\faCheck & {\cellcolor[rgb]{0.949,0.949,0.949}}Tag code execution       & {\cellcolor[rgb]{0.949,0.949,0.949}}Sandbox            \\ 
\midrule
\multirow{11}{*}{JavaScript} & Vue                                                    & \{\{ \}\}                                                           & \faCheck                                     & \faCheck                                     & Introspection                                                & -                                                      \\
                             & {\cellcolor[rgb]{0.949,0.949,0.949}}Pug                & {\cellcolor[rgb]{0.949,0.949,0.949}}\#\{ \}                         & {\cellcolor[rgb]{0.949,0.949,0.949}}\faCheck & {\cellcolor[rgb]{0.949,0.949,0.949}}\faCheck & {\cellcolor[rgb]{0.949,0.949,0.949}}Introspection            & {\cellcolor[rgb]{0.949,0.949,0.949}}-                  \\
                             & Handlebars                                             & \{\{ \}\} and \#                                                    & \faCheck                                     & \faCheck                                     & Introspection                                                & LCE                                                    \\
                             & {\cellcolor[rgb]{0.949,0.949,0.949}}\textbf{Marko}     & {\cellcolor[rgb]{0.949,0.949,0.949}}\$\{ \}                         & {\cellcolor[rgb]{0.949,0.949,0.949}}\faTimes & {\cellcolor[rgb]{0.949,0.949,0.949}}\faCheck & {\cellcolor[rgb]{0.949,0.949,0.949}}Introspection            & {\cellcolor[rgb]{0.949,0.949,0.949}}-                  \\
                             & Nunjucks                                               & \{\{ \}\}                                                           & \faCheck                                     & \faCheck                                     & Introspection                                                & -                                                      \\
                             & {\cellcolor[rgb]{0.949,0.949,0.949}}EJS                & {\cellcolor[rgb]{0.949,0.949,0.949}}\textless\%= \%\textgreater     & {\cellcolor[rgb]{0.949,0.949,0.949}}\faCheck & {\cellcolor[rgb]{0.949,0.949,0.949}}\faCheck & {\cellcolor[rgb]{0.949,0.949,0.949}}Introspection            & {\cellcolor[rgb]{0.949,0.949,0.949}}-                  \\
                             & doT                                                    & \{\{= \}\}                                                          & \faCheck                                     & \faCheck                                     & Introspection                                                & -                                                      \\
                             & {\cellcolor[rgb]{0.949,0.949,0.949}}Dust               & {\cellcolor[rgb]{0.949,0.949,0.949}}\{ \} and \{@ \}                & {\cellcolor[rgb]{0.949,0.949,0.949}}\faCheck & {\cellcolor[rgb]{0.949,0.949,0.949}}\faCheck & {\cellcolor[rgb]{0.949,0.949,0.949}}Bug                      & {\cellcolor[rgb]{0.949,0.949,0.949}}LCE                \\
                             & JsRender                                               & \{\{ \}\} and \{\{: \}\}                                            & \faCheck                                     & \faCheck                                     & Introspection                                                & -                                                      \\
                             & {\cellcolor[rgb]{0.949,0.949,0.949}}\textbf{Template7} & {\cellcolor[rgb]{0.949,0.949,0.949}}\{\{ \}\} and \{\{\# \}\}       & {\cellcolor[rgb]{0.949,0.949,0.949}}\faTimes & {\cellcolor[rgb]{0.949,0.949,0.949}}\faCheck & {\cellcolor[rgb]{0.949,0.949,0.949}}Tag code execution       & {\cellcolor[rgb]{0.949,0.949,0.949}}-                  \\
                             & \textbf{SquirrellyJS}                                  & \{\{ \}\} , @ and |                                                & \faTimes                                     & \faCheck                                     & Introspection                                                & -                                                      \\ 
\midrule
\multirow{5}{*}{Java}        & {\cellcolor[rgb]{0.949,0.949,0.949}}Thymeleaf          & {\cellcolor[rgb]{0.949,0.949,0.949}}\#\{ \}, \$\{ \} and [[ ]]      & {\cellcolor[rgb]{0.949,0.949,0.949}}\faCheck & {\cellcolor[rgb]{0.949,0.949,0.949}}\faCheck & {\cellcolor[rgb]{0.949,0.949,0.949}}Direct code execution    & {\cellcolor[rgb]{0.949,0.949,0.949}}No function calls  \\
                             & Pebble                                                 & \{\% \%\}                                                           & \faCheck                                     & \faCheck                                     & Introspection                                                & Sandbox                                                \\
                             & {\cellcolor[rgb]{0.949,0.949,0.949}}FreeMarker         & {\cellcolor[rgb]{0.949,0.949,0.949}}\$\{ \}                         & {\cellcolor[rgb]{0.949,0.949,0.949}}\faCheck & {\cellcolor[rgb]{0.949,0.949,0.949}}\faCheck & {\cellcolor[rgb]{0.949,0.949,0.949}}Functions code execution & {\cellcolor[rgb]{0.949,0.949,0.949}}-                  \\
                             & Jinjava                                                & \{\{ \}\}                                                           & \faCheck                                     & \faCheck                                     & Introspection                                                & Sandbox                                                \\
                             & {\cellcolor[rgb]{0.949,0.949,0.949}}Apache Velocity    & {\cellcolor[rgb]{0.949,0.949,0.949}}\$ and \#                       & {\cellcolor[rgb]{0.949,0.949,0.949}}\faCheck & {\cellcolor[rgb]{0.949,0.949,0.949}}\faCheck & {\cellcolor[rgb]{0.949,0.949,0.949}}Introspection            & {\cellcolor[rgb]{0.949,0.949,0.949}}-                  \\ 
\midrule
\multirow{2}{*}{Ruby}        & Slim                                                   & =                                                                   & \faCheck                                     & \faCheck                                     & Direct code execution                                        & -                                                      \\
                             & {\cellcolor[rgb]{0.949,0.949,0.949}}ERB                & {\cellcolor[rgb]{0.949,0.949,0.949}}\textless\%= \%\textgreater     & {\cellcolor[rgb]{0.949,0.949,0.949}}\faCheck & {\cellcolor[rgb]{0.949,0.949,0.949}}\faCheck & {\cellcolor[rgb]{0.949,0.949,0.949}}Direct code execution    & {\cellcolor[rgb]{0.949,0.949,0.949}}-                  \\ 
\midrule
Golang                       & Default engine                                         & \{\{ \}\}                                                           & \faCheck                                     & \faTimes                                     & -                                                            & LCE                                                    \\ 
\midrule
Perl                         & {\cellcolor[rgb]{0.949,0.949,0.949}}Mojolicious        & {\cellcolor[rgb]{0.949,0.949,0.949}}\textless\%= \%\textgreater     & {\cellcolor[rgb]{0.949,0.949,0.949}}\faCheck & {\cellcolor[rgb]{0.949,0.949,0.949}}\faCheck & {\cellcolor[rgb]{0.949,0.949,0.949}}Direct code execution    & {\cellcolor[rgb]{0.949,0.949,0.949}}-                  \\ 
\midrule
\multirow{2}{*}{.NET}        & ASP                                                    & \textless\%= \%\textgreater                                         & \faCheck                                     & \faCheck                                     & Direct code execution                                        & -                                                      \\
                             & {\cellcolor[rgb]{0.949,0.949,0.949}}Razor              & {\cellcolor[rgb]{0.949,0.949,0.949}}\@{ } and \@()                  & {\cellcolor[rgb]{0.949,0.949,0.949}}\faCheck & {\cellcolor[rgb]{0.949,0.949,0.949}}\faCheck & {\cellcolor[rgb]{0.949,0.949,0.949}}Direct code execution    & {\cellcolor[rgb]{0.949,0.949,0.949}}-                  \\
\bottomrule
\end{tabular}
\caption{This table shows the main characteristics of all the template engines analyzed in this work. \emph{A.A}. stands for Already analyzed, \emph{E.F.} for Exploit Found, \emph{LCE} for limited code execution. Template engine names in bold represent those we show as exploitable with RCE for the first time.}
\label{tab:templates-analyzed}
\end{table*}

Upon analyzing our results, we can draw conclusions regarding the current state of template engines. Figure~\ref{fig:templateOverview} presents aggregated findings, indicating that given the total number of engines analyzed, a rather low number of template engines integrate any form of protection. Of significant concern is also the fact that, among the 34 engines assessed, 31 allow or have allowed RCE. Notably, this susceptibility is more pronounced in the most popular programming languages, such as Python, PHP, Javascript, and Java.

A more detailed exploration of RCE types and security features is presented in Table~\ref{tab:templates-analyzed}. Here, we discern valuable insights into the kinds of RCE vulnerabilities and security mechanisms these engines possess. Introspective exploits are the most prevalent, affecting 16 template engines. Following closely are tags for code execution and direct code execution, impacting seven and six engines, respectively.

Regarding security measures, they are rarely implemented, with only ten out of the 34 engines offering any form of protection. Among these, sandboxes are the most commonly employed security feature. It is worth noting that prior works, as demonstrated in references~\cite{sstikettle, zhaoremote}, have raised doubts about the reliability of sandboxing. Out of the ten engines with security features, five incorporate a sandbox, while four provide limited code execution capabilities. This limitation on executable code within a template can significantly mitigate the risk of RCE but may also affect some of the engine's functionalities.

The results obtained from our analysis, along with what we saw in Section~\ref{sec:SSTIinTheWild:cves}, show that SSTI is still a vulnerability that is present in the wild, and its consequences can be critical. Despite this, the state of the art is not addressing the problem with new solutions due to the wide variety of template engines in different programming languages, which makes it difficult to develop solutions that can effectively mitigate SSTI risks. Furthermore, most developers might be unaware that selecting the proper template engine for certain web applications is essential. Instead, they might go with the most popular solutions or be unaware of the possibility that a template engine can be a security hazard for their applications. As Table \ref{tab:popularity} shows, the most popular templates also often allow RCE. Additionally, analyzing the scenario of CMSs, we can find insightful cases of SSTI in which the engine is directly accessible by users. CMS developers are responsible for ensuring that RCE is achievable in their product through SSTI. Moreover, by analyzing CVEs, we can see that most of them are critical because SSTI can be escalated to RCE. If the template engines used in those cases were not allowing RCE, the impact of SSTI would be much lower. This calls for a sensibilization among developers to select template engines considering worst-case scenarios in which their website or CMS is vulnerable to SSTI. Finally, the last prominent scenario is when template engines are provided as a service to platform users. We saw the example of GitHub pages, but many CMS allow users to use template engines to build email templates or web pages. In these cases, the risks and impact are even higher since the risk is to allow arbitrary users to take control of servers in which other instances of user content exist and potentially leak their private data or destroy other applications. In this scenario, selecting the proper template engine is crucial, but this calls for the automation of template engine analysis of RCE.

\subsection{Protecting against SSTI and RCE}
\label{sec:TemplateAnalysis:protecting}
SSTI vulnerabilities observed in the wild and reported in CVEs can generally be divided into two main categories, each stemming from different underlying issues.
\textbf{Constructing template strings from user input}. This scenario is particularly common in bug bounty reports and occurs when user input is embedded into a template string without proper sanitization. A typical example is when a user's registration data is included in a template string, allowing them to inject malicious instructions. To prevent SSTI in this context, developers should avoid constructing template strings that contain user-provided data whenever possible. If user-controlled data must be included within the template declaration, the input should be properly sanitized, specifically by removing or blocking sensitive template delimiters, such as curly brackets.
\textbf{Usage of template engines that allow RCE}. Most SSTI-related CVEs are assigned critical severity scores because these vulnerabilities often lead to remote code execution (RCE). However, selecting a framework or template engine that does not allow RCE can significantly reduce the severity and impact of an SSTI vulnerability. Using the data presented in Table~\ref{tab:templates-analyzed}, practitioners can quickly assess and choose template engines based on whether they permit RCE. It is crucial to emphasize that the selection of a secure template engine plays a key role in preventing RCE in the event that an SSTI vulnerability exists or if the application allows users to construct their own templates.

\section{Meaningful Examples}
\label{sec:examples}
In section~\ref{sec:teAndSSTI:overview}, we saw an example of template engine usage with \texttt{Jinja2}. This section will expand this overview by including other template engines and different programming languages. Notably, of the 34 template engines analyzed, nine were never analyzed before, and eight allow RCE. It would be interesting to provide a detailed explanation for each one but for space and repetitiveness reasons, we will only focus on those template engines that can provide useful insights into various coding practices, RCE types, and security features. For this reason, the only template engine that was never analyzed before that will be present in this section is Pyratemp (Section~\ref{sec:pyratemp}).
Nonetheless, we can briefly summarize our findings for the other eight engines that allow RCE and were never analyzed before.
\begin{itemize}
    \item \textbf{Cheetah}~\cite{cheetahdocs} is a \texttt{Python} template engine released in 2001. The documentation explains the existence of a special tag that can be used to execute direct \texttt{Python} code using \lstinline|{#echo <code>}|.
    \item \textbf{Genshi and Kid}~\cite{genshidocs} is a Python template engine born in 2006 that is based on another engine called \texttt{Kid}. Again, we can execute arbitrary Python code using special delimiters \lstinline|<?Python <code> ?>| .
    \item \textbf{web2py}~\cite{web2pydocs} is a \texttt{Python} framework inspired by the famous \texttt{Django}. In \texttt{web2py}, the engine allows access to introspective \texttt{Python} attributes like in \texttt{Jinja2}. Therefore, it is possible to use a payload similar to what we have seen in Section~\ref{sec:teAndSSTI:rceExample} to achieve RCE.
    \item \textbf{Chameleon}~\cite{chameleon_docs} is a \texttt{Python} engine that compiles its templates to optimize their execution speed. In this engine, we can quickly discover by trying to access the introspective attributes of an object that we can achieve RCE.
    \item \textbf{Laravel (Blade)}~\cite{laraveldocs} is an open-source framework to develop \texttt{PHP} web applications. It allows using different template engines, but by default, it uses its own, whose name is \texttt{Blade} \cite{bladedocs}. The documentation analysis reveals the presence of a raw \texttt{PHP} tag that allows code execution \lstinline|@php echo `ls` @endphp|.
    \item \textbf{Marko}~\cite{marko_docs} is a \texttt{JavaScript} engine focusing on simplicity and speed. Like many other \texttt{JavaScript} engines, it allows access to introspective object functions and attributes that can easily be exploited to achieve RCE. The following example shows what a \texttt{JavaScript} introspective payload looks like.
    \begin{lstlisting}[language=Javascript, caption=A JavaScript introspective payload for Marko engine., label=sec:examples:lst:markoPayload]
        ${ ''.toString.constructor.call({},"return global.process.mainModule.constructor._load('child_process').execSync('cat /etc/passwd').toString()")() }
    \end{lstlisting}
    This payload allows the execution of any shell command so a malicious user can exploit it to open a reverse shell and perform any operation on the server. The general idea is that we need to load the \texttt{child\_process} module and use it to call the \texttt{execSync} function that allows executing arbitrary shell commands in the server. The inner chain exploits the ability to traverse objects and move from a simple string to the global environment.
    \item \textbf{SquirrellyJS}~\cite{squirrelly_docs} is a \texttt{JavaScript} engine inspired by other famous names like \texttt{Nunjucks}, \texttt{Handlebars}, and \texttt{EJS}. By using the same exact payload above (Listing~\ref{sec:examples:lst:markoPayload})(with different delimiters), we can quickly achieve RCE with introspection.
    \item \textbf{Template7}~\cite{template7_docs} is a lightweight template engine focused on mobile-first web applications. In this case, looking at the documentation, we can find a special delimiter that allows us to execute arbitrary \texttt{Javascript} code. The following payload allows to execute arbitrary commands on \texttt{Template7}: \lstinline|{{js 'global.process.mainModule.require("child_process").execSync("ls")' }}|
\end{itemize}

These were the eight template engines that, through our analysis, we discovered to be allowing RCE. Now, we can move to five detailed examples of popular template engines in five programming languages (\texttt{Python}, \texttt{PHP}, \texttt{JavaScript}, \texttt{Java}, and \texttt{Ruby}). From these examples, we want to show insightful differences in how template engines address RCE and how they function in general.
Despite the presence of online blogs or cheatsheets of exploit payloads that concern the engines we analyzed, these examples aim to perform a more detailed analysis. Each subsection will show the main characteristics of a template engine, starting from simple safe and unsafe code examples and then diving into RCE paths, security features, and additional characteristics.

\subsection{Pyratemp: the Python Sandbox}
\label{sec:pyratemp}
\texttt{Pyratemp}~\cite{pyratempdocs} is a template engine focused on simplicity and speed. Its delimiters are the at symbol followed by an exclamation mark (\lstinline|@! <instruction> !@|). \texttt{Pyratemp} also provides a sandbox to protect against SSTI; we will analyze this sandbox in more detail after some examples of usage of this engine. To the best of our knowledge, no one has ever tried to analyze this template engine before.
\par The first example is for the safe usage of this template; the following code is not vulnerable to SSTI:
\begin{lstlisting}[language=Python, caption=Pyratemp safe code example., label=sec:pyratemp:lst:safeCode]
import pyratemp
user_input = request.form['username']
t = pyratemp.Template("Hello @!name!@.")
t(name=user_input)
\end{lstlisting}
In the code snippet above, we first import the \texttt{Pyratemp} engine module. Next, we retrieve user input from the HTTP request parameter \texttt{username}. Then, we create our template, supplying a string containing \texttt{Pyratemp} directives. In this case, we intend to display the value of the variable \texttt{name}. Finally, we execute the template by calling the \texttt{t} function (the name we assigned to the variable containing the template) and pass as many keyword arguments as needed in this function call. Each keyword argument represents a variable within the context of the template engine. In this instance, we pass the user input within the \texttt{name} variable.

The above code is safe from SSTI as we use the appropriate function to pass the user input in the template context. An example of code that instead is vulnerable to SSTI in \texttt{Pyratemp} is the following:

\begin{lstlisting}[language=Python, caption=Pyratemp vulnerable code example., label=sec:pyratemp:lst:vulnerableCode]
import pyratemp
user_input = request.form['username']
t = pyratemp.Template("Hello "+user_input+".")
t()
\end{lstlisting}

In this code snippet, we do not use the correct method to introduce user input into the engine. Instead of using a keyword argument when executing the template (the \texttt{t()} function call), we embed the user input directly within the \texttt{Pyratemp} code. This minor oversight means that the engine will execute any input from the user that includes the \texttt{Pyratemp} delimiters.

As we said in the introduction to this template engine, a pseudo-sandbox functionality checks the code we pass to the template. The sandbox is implemented with a class called \texttt{EvalPseudoSandbox}, allowing only a subset of \texttt{Python}'s built-ins that are considered safe by the engine creator. Additionally, it blocks the engine execution if the template contains any double underscore (e.g., \texttt{\_\_class\_\_}). As we saw previously with \texttt{Jinja2}, we can build RCE payloads using Python introspective attributes to call functions like \texttt{popen} or \texttt{system}. By filtering any template containing double underscores, this payload no longer works. To the best of our knowledge, there is no way to break out of this sandbox. Still, web developers should be careful not to pass dangerous objects to the template (e.g., objects with direct access to calls like \texttt{system} or other dangerous functions).

\subsection{Smarty: a PHP Sandbox}
\label{sec:examples:smarty}
\texttt{Smarty} \cite{smartydocs} is a template engine written in \texttt{PHP}. Its focus is on simplicity and security, and the syntax mainly uses single curly brackets (\lstinline|{ $variable }|). 

First, we can start with a safe example of its usage. This example is taken from the \texttt{Smarty} documentation and is divided into two files: a \texttt{smarty} source file that contains HTML tags and two variables from the template (\texttt{\$title\_text} and \texttt{\$body\_html}). The other file is the PHP source file, which compiles the template and sets the variable's value.

\par The first file we are going to see is the \texttt{index.tpl} file, which is HTML mixed with \texttt{Smarty} template code:
\begin{lstlisting}[language=HTML, caption=Smarty template code example., label=sec:smarty:lst:templateCode]
<!DOCTYPE html>
<html lang="en">
<head>
   <meta charset="utf-8">
   <title>{$title_text|escape}</title>
</head>
<body> {* This is a little comment that won't be visible in the HTML source *}
{$body_html}
</body> <!-- this is a little comment that will be seen in the HTML source -->
</html>
\end{lstlisting}
The above code contains some HTML tags and some \texttt{Smarty} directives. For example, we can see that the title tag contains a \texttt{Smarty} instruction delimited by the curly brackets, which displays the variable \texttt{\$title\_text} and applies the escape parameter to it. This operation will escape any HTML syntax in the variable. We also see a \texttt{Smarty} comment, which we can insert using the curly brackets and the asterisk symbol (\lstinline|{* comment *}|). Finally, we have the \texttt{\$body\_html} variable, which is not escaped because we want it to contain some HTML tags.
\par Now we can see the PHP code that renders this \texttt{Smarty} template:

\begin{lstlisting}[language=PHP, caption=Smarty safe code example., label=sec:smarty:lst:safeCode]
$smarty = new Smarty();
$smarty->assign('title_text', 'TITLE: This is the Smarty basic example ...');
$smarty->assign('body_html', '<p>BODY: This is the message set using assign()</p>');
$smarty->display('index.tpl');
\end{lstlisting}

The code shown above is shortened to the essential parts. Firstly, we create a new \texttt{Smarty} object representing the engine. Then, we can bind the variable in the engine using the assign function and pass the variable's name and the corresponding value. In this case, we set the variables we saw before \texttt{\$title\_text} and \texttt{\$body\_html}. Finally, we can render and display the template, which will become plain HTML (the engine will replace the variables when we call this function) and will be sent as a response to the user. It's important to emphasize that the above example, utilizing a separate template file (\texttt{index.tpl}), represents the conventional method for template rendering. In the preceding and subsequent examples, we will employ template strings for the sake of simplicity. However, creating a separate template file is the safest and most standardized approach to utilizing template engines. Nevertheless, there may be situations where web developers find it necessary to use template strings. In such cases, they must exercise utmost caution to prevent introducing SSTI vulnerabilities into their templates.

Now let us see an example of unsafe usage of the template instead; this time, we only report the \texttt{PHP} code for simplicity:

\begin{lstlisting}[language=PHP, caption=Smarty vulnerable code example., label=sec:smarty:lst:vulnerableCode]
$smarty = new Smarty();
$user_input = $_GET["username"];
$rendered = $smarty->fetch('Hi: '.$user_input);
\end{lstlisting}

The code is relatively short; as before, we must create our \texttt{Smarty} engine object and then fetch the user input using the global variable \texttt{\$\_GET}. Finally, we render the template code using the fetch function, putting the \texttt{Smarty} code directly as a string and concatenating the string "Hi: " to the user input. This practice is unsafe because the user input will be executed as part of the \texttt{Smarty} code, making this \texttt{PHP} application vulnerable to SSTI.

\par In \texttt{Smarty}, it is also quite easy to achieve RCE because we can exploit the existence of specific instructions that allow us to execute arbitrary \texttt{PHP} code, meaning that we can also call functions like \texttt{system} to execute arbitrary commands in the server. Here is an example of the usage of this kind of directive:

\begin{lstlisting}[language=PHP, caption=Smarty RCE payload using special delimiters., label=sec:smarty:lst:smartyPayloadSpecial]
{php}echo `ls`;{/php}
\end{lstlisting}

The \lstinline|{php}| directive indicates the start of a block that contains \texttt{PHP} code; in this case, we can use the backticks around a command we want to execute. They are an alias for executing system commands in \texttt{PHP}; then, we can print the command's output with echo.
This payload is not the only way to execute arbitrary \texttt{PHP} code. We can also abuse the standard \texttt{Smarty} syntax to execute system commands like in the following example.

\begin{lstlisting}[language=PHP, caption=Smarty RCE payload using regular delimiters., label=sec:smarty:lst:smartyPayloadRegular]
{system('ls')}
\end{lstlisting}

In this case, we are using the standard syntax (\lstinline|{}| ) and simply calling the system function, passing the command we want to execute in the server as an argument.

\par \texttt{Smarty} has a security option to set to true or false. By default, it is set to false, meaning there is no restriction, and the scenario we just explained is possible. When we set security to true, we instead have some restrictions. We can find them in \texttt{Smarty} documentation; the most important are:
\begin{itemize}
    \item The \texttt{php\_handling} setting is automatically set to \texttt{SMARTY\_PHP\_PASSTHRU}, which means that tags are echoed as-is. Therefore, nothing is executed as \texttt{PHP}, which instead happens with the default value \texttt{SMARTY\_PHP\_ALLOW}
    \item The \lstinline|{php} {/php}| tags are not allowed, eliminating the possibility of executing one of the previously seen payloads. 
    \item \texttt{PHP} functions are not allowed in if statements and as modifiers, except for the ones specified in the \texttt{security\_settings} variable. This disables the possibility to directly call the \texttt{system} function in templates.
    \item Templates and local files can only be included from certain directories specified in the \texttt{secure\_dir} variable
\end{itemize}
The activation of the above security features can help prevent RCE but it can still create problems. SSTI generally arises every time we allow users to use template syntax. Using these security features in the appropriate scenario means malicious users cannot exploit the vulnerability to perform an adversary server takeover. However, in the wrong scenario, it could still mean that an attacker can steal sensitive information. The right scenario would be where we want to allow the users to perform specific actions using the template engine. The wrong scenario is when programmers are not careful and treat the user input in unsafe ways because they know security measures are in place. The data passed to the template engine is still readable by attackers and could leak sensitive information. Furthermore, various sandbox bypasses for this template engine allowed RCE even when the security flag was enabled. Some have been fixed, but more might be found in the future.

\subsection{Dust: When Bugs Allow RCE}
\label{sec:examples:dust}
\texttt{Dust}~\cite{dustdocs} is an asynchronous template engine with over $17,000$ weekly downloads on npm. The delimiters are the single curly brackets (\lstinline|{<reference>}|), but there are special symbols that can be used after the opening curly bracket to use special functions (e.g. \lstinline|{@eq <comparison>}|).

\par We can now analyze this engine by seeing an example of safe code:
\begin{lstlisting}[language=Javascript, caption=Dust safe code example., label=sec:dust:lst:safeCode]
var dust = require('dustjs-linkedin');
var dust = require('dustjs-helpers');
var user = req.query.username
var compiled = dust.compile('<h1>Hi: {name}</h1>', 'test');
dust.loadSource(compiled);
dust.render('test', { name: user }, function(err, out) {
	html = out;
});
res.send(html)
\end{lstlisting}

In the first two lines, we import the \texttt{Dust} modules we need to render our template. Notice that we need to import \texttt{dustjs-helpers}. This detail is essential, as we need to use a helper function in this library to exploit this engine. In the third line, we collect the user input from the HTTP request. In the fourth line, we compile our template; the template prints "Hi: " followed by the user input safely passed through the \texttt{name} variable. In the \texttt{compile} function, we also set a name for our template, \texttt{test}, in this case. In the last lines, we load, render, and send our template; the \texttt{render} function takes three arguments. The first argument is the template's name (we set it before, and it is \texttt{test}). The second argument is the variables we need to use in the template, in this case, \texttt{name}, which contains the user input. The third argument is a function that either throws an error (\texttt{err} variable) or produces a string with the template (the \texttt{out} variable). If this function correctly runs, we end up with our rendered template in the \texttt{html} variable. Notice that this way of rendering the template differs from what we have seen for the other templates. This difference is because \texttt{Dust} uses asynchronous functions; the code is a bit longer, but performance is improved. In the last line, we send the generated template to the client.

The above code is safe from SSTI because we are not using the user input directly inside the template code but passing it using the proper argument in the render function. The following code instead does not treat user input safely and is vulnerable to SSTI.

\begin{lstlisting}[language=Javascript, caption=Dust vulnerable code example., label=sec:dust:lst:vulnerableCode]
var dust = require('dustjs-linkedin');
var dust = require('dustjs-helpers');
var user = req.query.username
var compiled = dust.compile('<h1>Hi: '+user+'</h1>', 'test');
dust.loadSource(compiled);
dust.render('test', {}, function(err, out) {
	html = out;
});
res.send(html)
\end{lstlisting}

The main difference in the above code with respect to the safe one is in the fourth and sixth lines. In the fourth line, we concatenate the user input to the template code, which is why this code is vulnerable to SSTI. In the sixth line, we pass an empty object where we should pass the user input as we did in the safe code.

\par The following payload can be used to exploit SSTI to achieve RCE in \texttt{Dust} with a version of the \texttt{dustjs-helpers} module before or equal to \texttt{1.5.0}. In fact, in the versions after this, the \texttt{if} helper was removed for security reasons, as it allowed for arbitrary \texttt{JavaScript} code to be evaluated.

\begin{lstlisting}[language=Javascript, caption=Dust payload for RCE using if helper eval bug., label=sec:dust:lst:payload]
{@if cond="eval('global.process.mainModule.require(\'child_process\').execSync(\'curl  https://evil.com/?res=`ls`\').toString()')"}{/if}
\end{lstlisting}

The above exploit shows that we can inject an \texttt{eval} in the condition by using the if statement. Inside the \texttt{eval}, an attacker can execute a \texttt{JavaScript} introspective payload to perform arbitrary system commands. The introspective payload aims at importing the \texttt{child\_process} module and calling the function \texttt{execSync}. In this case, the unique payload we crafted for this example does not simply execute \texttt{ls}, but it executes \texttt{curl}. The reason is that since we are inside an if condition, the returned value of \texttt{eval} is converted to a boolean, and the output is discarded. To exploit this payload to achieve RCE, an attacker can either open a reverse shell or send the command output to an external server.

\subsection{Jinjava: Java Introspection to RCE}
\label{sec:examples:jinjava}
\texttt{Jinjava} is a template engine based on Python's \texttt{Jinja2}, as the name suggests. The syntax of \texttt{Jinjava} is very similar to the one we saw for \texttt{Jinja2} (it uses \lstinline|{{ }}| as delimiters), and the vulnerabilities are similar with the exception of the programming language. We will see that the developers of this template engine patched it to avoid allowing RCE. Before diving into this template's vulnerable side, let us start by seeing an example of the safe usage of this template engine.
\begin{lstlisting}[language=Java, caption=Jinjava safe code example., label=sec:jinjava:lst:safeCode]
String user = request.getParameter("username");
Jinjava jinjava = new Jinjava();
Map<String, Object> context = Maps.newHashMap();
context.put("name", user);
String renderedTemplate = jinjava.render("Hi: {{ name }}", context);
\end{lstlisting}
Firstly, we fetch the user input using the \texttt{getParameter} function. Then, we instantiate a \texttt{Jinjava} object and will use it to render the template code. Then, we create a \texttt{hashmap} object. It will contain the context for our template engine, and we can put the internal variables we need in the engine context. We bind the user input to a template variable called \texttt{name}. Finally, we use the render function to parse our \texttt{Jinjava} code that should display the message "Hi: " followed by the user's input. This procedure is safe as the user input will not be directly parsed by the template engine but will only be substituted once we render the template. Now let us see instead an example of unsafe usage of the template engine.
\begin{lstlisting}[language=Java, caption=Jinjava vulnerable code example., label=sec:jinjava:lst:vulnerableCode]
String user = request.getParameter("username");
Jinjava jinjava = new Jinjava();
Map<String, Object> context = Maps.newHashMap();
String renderedTemplate = jinjava.render("Hi: "+user, context);
\end{lstlisting}
The code is similar to the previous one; the context is empty in this case. Instead of safely binding the user input to a template variable, we are directly concatenating it in the \texttt{Jinjava} code. This mistake can cause an SSTI vulnerability, but how serious is it in this engine? Can we achieve RCE?
\texttt{Jinjava} developers put some effort into trying to avoid RCE in their template. Also, a newer version of JDK helped in this intent by removing a particular functionality that was being used in exploits. Before explaining in more detail what has changed, let us analyze the payload that works on older versions of \texttt{Jinjava} and the JDK.

\begin{lstlisting}[language=Java, caption=Jinjava introspective payload example., label=sec:jinjava:lst:payload]
{{'a'.getClass().forName('javax.script.ScriptEngineManager').newInstance().getEngineByName('JavaScript').eval('var x=new java.lang.ProcessBuilder; x.command("whoami")'); x.start()')}}
\end{lstlisting}
The payload is not as simple as the ones we saw in Python's \texttt{Jinja2}, but the rationale is the same: we exploit the introspective nature of object-oriented programming. We start with a string \texttt{'a'} and call the \texttt{getClass} method, which allows us to access any class instance by calling the \texttt{forName} method. We get an instance of \texttt{ScriptEngineManager}, a particular class we can use to execute scripts in other languages like, in this case, \texttt{JavaScript}. The \texttt{eval} method is the final piece that allows executing Javascript code to execute arbitrary commands and achieve RCE. Luckily for us, this payload will not work on newer versions of \texttt{Jinjava}; this is because, being an open-source project, someone opened an issue on GitHub, underlining that we can prevent this payload from working if the \texttt{getClass} function is not callable. Hence, the developers restricted this function, which is now not callable on the template code. Also, if we use a recent version of JDK (e.g., JDK 18), this payload will not work because the \texttt{ScriptEngineManager} class does not have the \texttt{Javascript} engine by default, so an attacker cannot use this trick to achieve RCE.

\subsection{ERB: Ruby Code Execution}
\label{sec:examples:erb}
\texttt{ERB} is a \texttt{Ruby} template engine in which the syntax is based on angular parenthesis (\lstinline|<%= instruction %>|). As we saw in other engines, SSTI arises if user inputs are not treated securely. The code below is an example of the safe usage of this template engine.

\begin{lstlisting}[language=Ruby, caption=ERB safe code example., label=sec:erb:lst:safeCode]
class Env
  attr_accessor :name
end

on param("username") do user
scope = Env.new
scope.name = user
template = Tilt['erb'].new() {|x| "<%= name %>"}
res.write template.render(scope)
\end{lstlisting}
 
In this example, our first step is to define a class, \texttt{Env}, which serves as a container for organizing the data we intend to incorporate into the template. This class possesses a parameter \texttt{name}. Following this class declaration, we proceed to collect the user input received by the server, \texttt{username} in this case. Subsequently, we will refer to this input with the user variable.
Now, we create an instance of the \texttt{Env} class by invoking the \texttt{Env.new} method, and we assign this instance to a variable named \texttt{scope}. To complete the setup, we set the \texttt{user} attribute of the \texttt{scope} object to the value stored in the \texttt{user} variable.
The next step involves crafting our template, which is facilitated by the \texttt{Tilt} module. This module enables us to choose from a variety of templates. In this specific case, we access the \texttt{erb} template and instantiate it using the \texttt{new} function, passing the desired parameters. In this instance, the parameter \texttt{x} represents the string we wish to render as a template.
Lastly, our final instruction entails writing the content returned by invoking the \texttt{render} function on our previously created \texttt{template} variable. It is worth noting that while \texttt{Ruby} employs a distinct syntax compared to other programming languages we have examined, the underlying mechanism remains consistent: we acquire and bind user input to a variable that is subsequently passed to the template. Once the template is rendered, the webpage effectively displays the variable's value, ensuring a safe and secure presentation of data.

\par An example of unsafe usage of the template is instead provided in the following code:
\begin{lstlisting}[language=Ruby, caption=ERB vulnerable code example., label=sec:erb:lst:vulnerableCode]
on param("username") do user
template = Tilt['erb'].new() {|x| user}
res.write template.render
\end{lstlisting}

The general idea is the same, but in this case, we do not provide any environment to the template from which it can present the user input inside the page; in this case, we render the user input. On the first line, we fetch it from the request's parameters. Then, we create the template by passing the user input to it and rendering it, making it very easy for an attacker to inject template syntax that will be executed and rendered in the output.

In this scenario, it is very easy for the attacker to perform an SSTI, allowing arbitrary code execution. \texttt{ERB} tags can execute arbitrary \texttt{Ruby} code, so we can simply execute the following payload:
\begin{lstlisting}[language=Ruby, caption=ERB RCE payload with direct code execution., label=sec:erb:lst:payload]
<%= IO.popen('ls /').readlines()  %>
\end{lstlisting}

In this case, the command executed is \texttt{ls}, but any command can be executed. The \texttt{readlines} function allows the attacker to retrieve the command output.

\section{Lessons Learned}
\label{sec:lessons_learned}

In this section, we summarize the main insights that resulted from our analysis, highlighting which pitfalls are still present in SSTI research.

\begin{enumerate}
    \item \textbf{Existing defenses against SSTI are scarce.}
    We have demonstrated the widespread use of template engines in real-world applications and the persistent nature of SSTI vulnerabilities. Numerous CVEs related to SSTI in applications underscore their significant impact on security. We have shown that the severity of SSTI is reflected both in high CVE scores and in the substantial payouts offered by bug bounty programs. Nevertheless, the issue persists, and neither existing tools nor current mitigation efforts have proven sufficient.
    
    \item \textbf{RCE in template engines is common and inadequately prevented.}
    We conducted an in-depth analysis of 34 template engines, including the first-ever examination of nine engines—eight of which allowed RCE. From this analysis, we extracted valuable insights, identifying four distinct types of RCE paths and categorizing four possible protection approaches against RCE attacks. This study emphasizes the need for automated, cross-language methods to determine whether a template engine allows RCE. Since 30 out of the 34 analyzed template engines have exhibited RCE paths—21 of which still do—we conclude that RCE is a prevalent issue. We also engaged in a comprehensive discussion on why RCE vulnerabilities persist in template engines and how sandboxing mechanisms, the most common line of defense, can be bypassed.
    
    \item \textbf{SSTI is underestimated, and research efforts to prevent this vulnerability are lacking.}
    We have shown that the only defense analyzed in academic research is instruction set randomization, which is difficult to implement in practice. Furthermore, the limited research on this topic has largely focused on SSTI and RCE detection techniques rather than exploring innovative methods for prevention.
\end{enumerate}

\section{Conclusions}
\label{sec:conclusion}
In this article, we have thoroughly examined template engines from multiple angles. To start, we explained their general functioning, both in theory and practical application. Following that, we delved into the vulnerabilities associated with their usage, specifically investigating SSTI and RCE pathways. We also scrutinized the existing tools and research related to SSTI. Furthermore, we offered a detailed exposition of our analysis findings, including case studies and practical examples. Our objective in this endeavor was to bring attention to the RCE problem in this widely utilized technology in modern web applications. By showing the repercussions of SSTI and the limitations of existing research efforts, we aim to inspire future works that address the research needs of template engines. The analysis of current and past works has shown that the efforts on SSTI have been focused on the exploitation, detection, and sandbox evasion parts, whilst the problem of RCE remains mostly unexplored. We believe that future works should focus on this issue, finding ways to build effective defenses against RCE paths in template engines. By focusing on mitigating RCE, the overall impact of SSTI vulnerabilities will decrease noticeably.

\begin{acks}
This work was partially supported by project SERICS (PE00000014) under the NRRP MUR program funded by the EU - NGEU.
\end{acks}

\bibliographystyle{acm}
\bibliography{software}


\end{document}